%% file: H4172.tex
\documentclass{aa}

\usepackage{graphics,rotating,epsfig}
\usepackage{natbib}

\input{H4172mac}
\bibpunct{(}{)}{;}{a}{}{,}

\newcommand{\halfa}{H$\alpha$}
\newcommand{\iras}{IRAS\,08544 }
\begin{document}


\title{IRAS\,08544-4431 : a new post-AGB star in a binary system surrounded
by a dusty disc.
\thanks{based on observations collected at the European Southern Observatory
in Chile (62.L-0508) and at SAAO. The radial velocity data was obtained with
the Swiss 1.2m Euler telescope at La Silla, Chile.}}

\author{Thomas Maas\inst{1}
\and Hans Van Winckel\inst{1}\thanks{Postdoctoral fellow of the Fund
for Scientific Research, Flanders}
\and Tom Lloyd Evans\inst{2}
\and Lars-\AA ke Nyman\inst{3,4}
\and Dave Kilkenny\inst{5}
\and Peter Martinez\inst{5}
\and Fred Marang\inst{5}
\and Francois van Wyk\inst{5}}

\offprints{Thomas Maas}
\mail{Thomas.Maas@ster.kuleuven.ac.be}
\institute{Instituut voor Sterrenkunde, K.U.Leuven, Celestijnenlaan 200B,
B-3001 Leuven, Belgium \and
School of Physics and Astronomy, University of St Andrews, North Haugh,
St Andrews, Fife, Scotland KY16 9SS \and
European Southern Observatory, Casilla 19001, Santiago 19, Chile
\and Onsala Space Observatory, S-43992 Onsala, Sweden
\and South African Astronomical Observatory, P. O. Box 9,
Observatory 7935, South Africa}
\date{Received  / Accepted}


\abstract{We present an analysis of our extensive data-set on
 IRAS\,08544-4431. It is the first object we discuss of our newly defined 
sample of stars, selected for their position in the `RV\,Tauri' box in the IRAS
$[12]-[25]$, $[25]-[60]$ two-color diagram. Moreover, our selection criteria
included an observed excess in the L-band, indicative of a dusty disc.
 The SED of IRAS\,08544-4431 shows a broad IR excess starting already at $H$. 
Our optical photometric data reveal some evidence for  deep and
shallow minima in the light curve and a pulsation time-scale of around 100 days
 with a small amplitude ($\Delta V$ peak-to-peak = 0.17 mag).
Our CORALIE radial velocity measurements show that IRAS\,08544-4431 is a
binary system with a period of  499 $\pm$ 3 days and a mass function of
0.02 M$_{\odot}$. Moreover, IRAS\,08544-4431 is
detected in both the CO (2-1) and (1-0) mm-wave emission lines.
The triangular shape of the weak CO profile confirms  that
part of the circumstellar  material is not freely expanding but resides
  probably in a dusty circumbinary disc.
 Our  chemical abundance analysis of
a high resolution spectrum of high S/N reveals that a depletion process has
modified the photospheric abundances to a moderate extent ([Zn/Fe]=+0.4).
All these findings confirm that the F-type IRAS\,08544-4431 is another
 good example
of a binary Post-AGB star surrounded by a dusty disc. The \halfa\,
P-Cygni profile shows ongoing mass-loss with a very high outflow 
velocity, the origin of which is not understood. 
The strength and velocity of the \halfa-absorption are modulated 
with the orbital motion; the maxima of both quantities 
($\sim$ 400 \kms, 5 \AA\, respectively) occur at superior conjunction. 
}

\titlerunning{IRAS\,08544-4431}
\authorrunning{T. Maas et al.}
\maketitle


\section{Introduction}

A new set of probable post-AGB stars was found in the IRAS Point Source
Catalog by Lloyd Evans (1999 and in preparation). Their recognition follows
the finding that the brightest of the RV\,Tauri stars contained in the
General Catalogue of Variable Stars have IRAS photometry which includes
a reliable 60\,\mic\ measurement, so that they may be plotted in a
 $[12]-[25]$, $[25]-[60]$ two-color diagram. \citet{1985MNRAS.217..493E}
 and \citet{1989MNRAS.238..945R} remarked on the fact
 that the region containing
these data lies in a well-defined and relatively thinly-populated part of
the diagram, which corresponds to the greater relative brightness of these
stars at 60\,\mic\ compared to the majority of the M-type variables which have
extensive dust shells. This made feasible a search of the IRAS photometry for
new examples of RV\,Tauri stars, selected by their mid-infrared colors
rather than by their variability in blue light. A list of candidates was
prepared by calculating the colors of all IRAS stars with reliable
(quality 3) photometry at 60\,\mic\ and selecting those which were located in
a rectangle in the IRAS two-color diagram which contained most of the known
RV\,Tauri stars. \\
\\
 The defining rectangle is $$
 \left\{\begin{array}{ll} [12]-[25] \equiv & 1.56 + 2.5 \log[F(25)/F(12)] = 1.0 - 1.55  \\ \, [25]-[60] \equiv & 1.88 + 2.5 \log[F(60)/F(25)] = 0.20 - 1.0 
  \end{array} \right. $$ 
\\
\\
Over 1000 sources south of 20\,deg N declination were located on the Palomar,
ESO and/or SRC Sky Survey prints. Those where a reasonably bright star was
found were selected for visible-light spectroscopy and near infrared ($JHKL$)
photometry. Initial spectroscopic observations made with the Unit
Spectrograph and RPCS detector system at the Cassegrain focus of the 1.9m
Radcliffe reflector at the Sutherland observing site of the South African
Astronomical Observatory covered the approximate range
3500 -- 6700 \AA, with a resolution of 7 \AA. This survey revealed stars
of a wide range of spectral types with the excess at L indicative of the
presence of a dusty disk \citep{1997Ap&SS.251..239L}.
 Those objects with spectral
types, based primarily on the hydrogen lines in view of the many spectral
peculiarities of RV\,Tauri star spectra, of F, G or K were regarded as
possible RV\,Tauri stars and were selected for additional observations.
These included $UBVRI$ photometry, of those stars which were not too faint
and did not suffer crowding problems, made at Sutherland as part of the SAAO
service observing programme.

\begin{table}
\caption{Basic parameters of \iras (The coordinates are from the GSC).}
\label{tab:basicdata}
\begin{center}
\begin{tabular}{lrr} \hline
      &       & \iras  \\
\hline
Coordinates & $\alpha_{2000}$ & 08 56 14.2  \\
            &  $\delta_{2000}$ & -44 43 11.0 \\
Galactic & l & 265.51  \\
 coordinates  & b & +0.39  \\
\hline
\end{tabular}
\end{center}
\end{table}

Repeated near infrared and/or visible light photometry showed that while some
of the RV\,Tauri-like stars are variables of substantial amplitude, most are
less variable than typical RV\,Tauri stars. \cite{1999IAUS..191..453E}
 reported a
correlation between amplitude and spectral type: the most variable stars
have spectral types in the narrow range occupied by the bulk of the
previously-known RV\,Tauri stars, while those of lesser amplitude are mainly
of earlier F subtype. This suggested that many IRAS detected
 RV\,Tauri stars of the GCVS are
simply those stars with dusty disks which happen to be passing through the
instability strip of the Type II Cepheids \citep{1999IAUS..191..453E}.

We embarked on monitoring programs both in optical and near-IR photometry at
SAAO and radial velocity at La Silla. These data were supplemented with
single observations to obtain high-resolution spectroscopy
and CO mm-wave line emission. The aim is to characterize these evolved
objects, compare them with GCVS RV\,Tauri stars and to discuss the
 relationship between the presence of such a disc and the eventual binarity of
the central object.

In this first paper devoted to this project, we report on the analysis of an
extensive data-set obtained for one object: IRAS\,08544-4431 (\iras in what
 follows, see Table~\ref{tab:basicdata}).
It is one of the stars in our sample with a smaller amplitude and
 is of spectral type F3, which was determined on the basis of a 
low resolution spectrum \citep{1999IAUS..191..453E}. The object was
selected, despite the low quality flag (2) of the 60 $\mu$m flux point, because
of its LRS spectrum.
 In the literature this little studied object
is referred to as a carbon star following the LRS IRAS spectral classification
\citep{1997ApJS..112..557K}. The LRS spectrum is, however, rather featureless
and in this paper we argue that the spectrum is not that of a C-rich
object \citep{1986A&AS...65..607O} and that the star is certainly not a
carbon star.
In following papers we will describe the overall characteristics of the sample.
 In  Sect.2 we outline the observational dataset and the reduction
 procedures, next we discuss the Spectral Energy Distribution (SED). In Sect.4
we show that \iras is variable with a small amplitude and long timescale.
There is some hint that the defining light curve of GCVS RV\,Tauri stars is 
also present in \iras, but a dedicated monitoring program is needed to study 
this in more detail.
In Sect.5 the radial velocity measurements are presented, proving that \iras
is a binary; we also used our CORALIE data
to study the variation of the P\,Cygni profile of \halfa. In Sect. 6 we present
our CO line emission measurements and in Sect.7  our chemical abundance
 analysis shows that depletion, which is a widespread phenomenon among
RV\,Tauri stars, has slightly altered the photospheric abundances of \iras 
as well.
In Sect.8  we put our main conclusions together and end with a discussion.


\section{Observations}

\subsection{Photometry}

Infrared observations on the $JHKL$ system defined by the standard stars
of \cite{1990MNRAS.242....1C} were made with the Mark II
Infrared Photometer on the 0.75\,m reflector at Sutherland (SAAO
Facilities Manual). $L$ has the mean
effective wavelength of 3.5\,\mic\ and not 3.8\,\mic\ for which the same
symbol is sometimes used. Standard errors for this bright source are
 0.03\,mag for $J$, $H$ and $K$ and 0.05\,mag for $L$. Individual observations will be 
given along with those of other stars by Lloyd Evans (in preparation).

Photometric observations on the $UBVRI$ system were made with the Modular
Photometer on the 0.5\,m telescope at Sutherland on 54 nights from JD\,2450858
to 2451666, with repeated observations on several nights. Observations were
continued with the 0.75\,m Automatic Photoelectric Telescope (APT)
\citep{2002MNSSA..61..102M} on 37
nights from JD\,2451929 to 2452060. This concentrated run of observations
made with the APT proved invaluable in showing that the pulsation time-scale
 was of the order of 100 days. The observations were made with reference 
to the Cousins
$UBVRI$ standards in the E-regions \citep{1989SAAOC..13....1M}, and
reduced to the
standard system using color terms derived from observations of standard
stars of a wide range of color. There was no local standard star and hence
the data include the increased random errors involved in a magnitude transfer
through a large angle. The nominal error of a single observation is typically
0.007\,mag but may be substantially larger in $U-B$, especially on bright
moonlit nights, to which has to be added the contribution, perhaps of order
0.01\,mag, of the E-region transfer.

\subsection{Spectroscopy}

\subsubsection{Low Resolution spectra.}

 The first spectra were taken with the RPCS detector on the Unit Spectrograph 
at the Cassegrain focus of the 1.9\,m Radcliffe reflector at
the Sutherland observatory of the SAAO. Spectra taken over the range
3500-7000\,\AA\ with a resolution of 7\,\AA\ were classified against a set
of MK standard star spectra and a hydrogen-line type of F3 was obtained
from the higher Balmer lines (since \halfa\, at least is always in
emission) in the same way as for a large sample of known RV Tau stars
observed earlier with a photographic detector on the same
spectrograph. \iras is thus one of the stars, described
as RV Tau-like on the basis of their spectral energy distribution and F-G
spectral type, which lie outside and on the hot side of the instability
strip (Lloyd Evans 1999).

\subsubsection{FEROS spectrum}
A high signal-to-noise and high resolution ($\lambda/\delta\lambda \sim 48000$)
 spectrum was obtained with FEROS mounted on the 1.52 m ESO telescope at La
 Silla (Chile). FEROS is a fiber fed echelle spectrograph which  covers the
 complete optical region from 3700 to 8600 \AA.
The spectrograph is equipped with a CCD camera incorporating a chip with
2048 x 4096 pixels of 15 x 15 $\mu$m.
The online reduction consists of
 background subtraction, order extraction, flatfielding and wavelength
 calibration, using a Th-Ar spectrum.
We normalized the reduced spectra using smoothed splines on interactively
determined continuum points. The FEROS spectrum
 was obtained on January 26th 1999. Sample spectra are shown in
Figs.~\ref{spec6740} and~\ref{spec5660}.

\subsubsection{CORALIE data}
We have been monitoring \iras with the spectrograph CORALIE mounted on the
Swiss telescope at La Silla, ESO Chile. CORALIE \citep{1999psrv.conf...13Q}
 is a high resolution ($\lambda/\delta\lambda = 50000$) fiber
fed echelle spectrograph which is constructed to perform high precision
velocity measurements. Cross-correlation profiles are calculated on-line with a
standard F mask. To improve the cross-correlation profiles we
constructed a individual mask for \iras tuned on the basis of the FEROS spectrum.
The radial velocities are obtained by Gaussian fitting of these profiles.
We started this monitoring in November 1998 and have gathered 107 data
points over a total time span of 1435 days.

\subsubsection{CO measurements}
CO measurements were performed with the 15 m Swedish-ESO submillimeter
 telescope (SEST) at La Silla. A dual channel SIS receiver was used
to observe simultaneously at 115 GHz (the J = 1-0 line) and at 230 GHz
(the J = 2-1 line). At 115 GHz the half power beam width and the main beam
 efficiency of the telescope are 45\arcsec\, and  0.7  and at 230 GHz
 23\arcsec\, and 0.5.
 Typical receiver temperatures are 110 K (115 GHz) and 150 K (230 GHz).
 Three Acousto Optical Spectrometers (AOS) were used as backends.
The 230 GHz receiver was connected to a high resolution spectrometer (HRS) and
 a low resolution spectrometer (LR1), the 115 GHz receiver to a second low
resolution spectrometer (LR2). Five exposures were taken in the period from
October 1999 to August 2001.
The system temperatures for the CO (2-1)
observations varied between 250 K and 350 K. The intensity scales, reported
in Fig.~\ref{iras08544co} and Table~\ref{tabco}, are given
as main beam brightness temperature,
 $T_{MB} = T_{A}^{\ast} / \eta_{MB}$,
 where $T_{A}^{\ast}$ is the antenna temperature corrected for atmospheric
attenuation using the chopper wheel method
and $\eta_{MB}$ is the main beam efficiency.

Additionally we mapped the environment of \iras with low resolution
 spectra of (J = 2-1) CO on July 28th 2000. A map spacing of 11\arcsec\,
 was used.
Finally, also one low resolution (J = 1-0) HCN spectrum was taken. No line was
 detected to an r.m.s. level of 0.003 in $T_{MB}$.
For all observations we used dual beam switching, where the source is
alternately placed in the signal and the reference beam, using a beam throw
of about 12\arcmin\,.


\section{Spectral Energy Distribution (SED)}

\begin{figure}
\resizebox{\hsize}{!}{\includegraphics{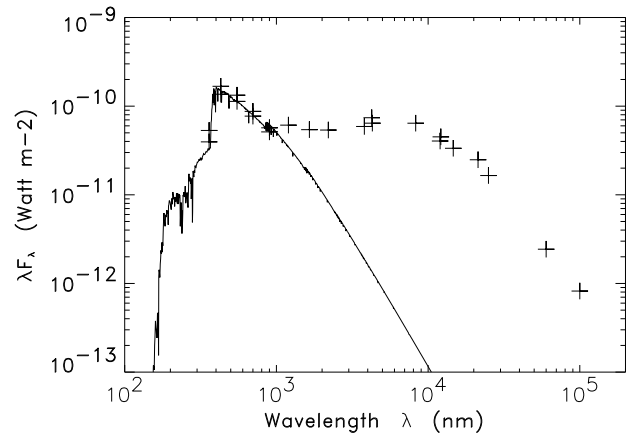}}
\caption{\label{sed} The SED of \iras. For the optical measurements,
 the minimal and maximal fluxes are displayed. The photometric data
are given in Table~\ref{tab:tabsed}. The dereddened
 fluxes are shown using a total reddening of $E(B-V)$=1.45 with the
 standard reddening law.
Note the broad IR excess starting already at $H$.}
\end{figure}

\begin{table}
\begin{center}
\caption{ The photometric data used to construct the SED. For the optical and 
near-IR photometry, the mean magnitude, the standard variation ($\sigma$) and 
the number of observations (N) is shown. The IRAS and MSX fluxes are given in 
Jansky.  The IRAS quality label and for MSX the percent error is shown.}
\label{tab:tabsed}
\begin{tabular}{l|rrr||r|rr}
 & $<$m$>$ & $\sigma$ & N & $\lambda$ & flux & error \\
\hline
U & 11.85 & 0.06 & 103 & 12 \mic  & 180.30 &  \\      
B & 10.31 & 0.05  & 103 & 25 \mic & 158.80 &  \\      
V & 9.13 & 0.03 & 103 &   60 \mic & 56.25 & :  \\     
R$_{C}$ & 8.14 & 0.03 & 103 &   100 \mic & 28.43 & L  \\    
I$_{C}$ & 7.11 & 0.03 & 103 &   B1 (4.29 \mic) & 80.17 & 9.7  \\  
J & 5.65 & 0.02 & 2 &	     B2 (4.25 \mic)  & 91.45 & 9.1  \\  
H & 4.67 & 0.02 & 2 &	     A (8.28 \mic) & 163.09 & 5  \\    
K & 3.51 & 0.02 & 2 &	     C (12.13 \mic) & 173.50 & 3  \\     
L & 1.59 & 0.02 & 2 &      D (14.65 \mic) & 156.70 & 4  \\     
&        &      &   &      E (21.34 \mic) & 171.57 & 6  \\     
\end{tabular} 
\end{center}
\end{table}

 All photometric data used in constructing the SED are given in
 Table~\ref{tab:tabsed}. The optical coordinates of the Guide Star Catalogue
 optical object is some 19\arcsec\, west, and 7\arcsec\, north of the IRAS
 position and on the edge of the 95$\%$ positional uncertainty ellips as given
 in the IRAS point source catalog.
The near-IR images obtained with the PICNIC camera \citep{1999MNSSA..58..147G}
 in $J$, $H$ and $K$ did not show other near-IR sources which might
 be connected with the IRAS source. Moreover, the MSX 
\citep{1999STIN...0014854E} position shows a correct association with the
 optical source so we can safely assume the IRAS fluxes are associated
 with the optical component as well.

The spectral energy distribution is shown in Fig.~\ref{sed}.
The full line represents the Kurucz model with the parameters obtained
from analysis of the FEROS spectrum (see Sect.7).
The total reddening of $E(B-V)$=1.45 is found by minimizing the difference
 between the dereddened fluxes in the optical and in the J-band and the model atmosphere.
A rough estimate of the interstellar component of the reddening can be obtained
using the relation between the DIB strength at 5780 and 5797 \AA\ and the
reddening \citep{1999A&A...347..235K}. There is a rather
large scatter in this relation
(correlation coefficient 0.8-0.89) but we obtain a $E_{IS}(B-V)$=0.7$\pm$0.2.

With [12]-[25] = 1.42 and  
[25]-[60]=0.75 \iras falls clearly into the 'RV\,Tauri' box of the colour-colour
diagram which is shown in Fig. 3(c) of \citet{1985MNRAS.217..493E}. Moreover,
\iras shows a very broad IR excess, which starts already
 at $H$ (1.65 $\mu$m) and points to
 the presence of hot dust in the close environment of the star.
Also this hot circumstellar dust component is an observed feature
for RV\,Tauri stars \citep{1972ApJ...178..715G,1972PASP...84..768G,
1985MNRAS.217..493E}, although
compared to the RV\,Tauri stars studied in the last reference, the excess in
$H$, $K$ and $L$ for \iras is large.

This near-IR excess is generally not observed in single post-AGB stars
 with an expanding dust shell. It is, however, not unique and several post-AGB
stars also show a similar SED with a broad-IR excess.
 All these latter objects turned 
out to be binaries and for these objects the near-IR excess is interpreted as
 coming from hot dust which is stored in a dusty disc. The indirect and direct 
evidence for this interpretation is given in \citet{2001Ap&SS.275..159V} and 
references therein. For several objects, the disc is resolved (e.g. AC\,Her, 
\citealt{2000ApJ...541..264J}).

We conclude that the SED of \iras shows a very broad IR excess around a 
weakly variable F-type central star, which indicates the presence of
a circumstellar disc instead of a freely expanding outflow. Its IR colours are
 very similar to those of the bulk IRAS detected RV\,Tauri stars.


\section{{\bf Pulsations}}

\begin{figure}
\resizebox{\hsize}{!}{{\includegraphics{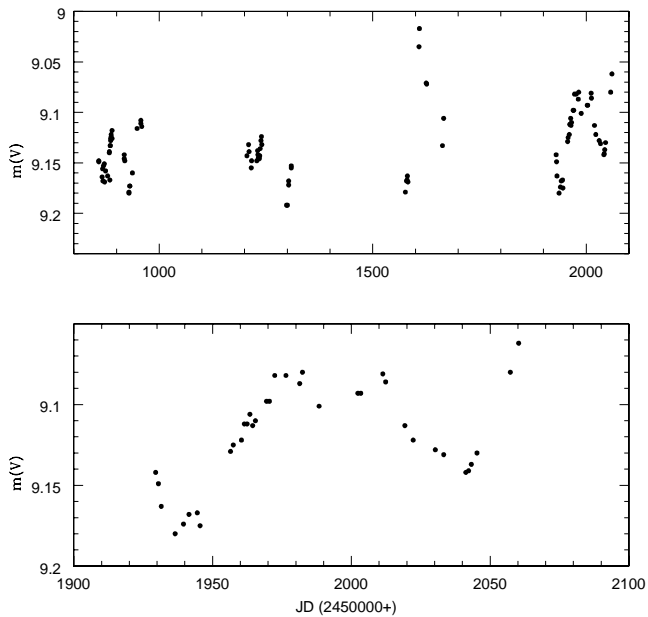}}}
\caption{\label{phot}
The V-band photometric variations displayed by IRAS\,08544 : upper panel :
all photometry, lower panel : only the APT photometry.
 The amplitude is low ($\Delta V$ peak-to-peak = 0.17 mag) and
 the pulsation time-scale  is near 100 days.}
\end{figure}

\iras is clearly variable, albeit with a small peak-to-peak amplitude
of 0.17 mag in $V$ and 0.23 mag in $B$. Since \iras has the same IR properties
as RV\,Tauri stars, we checked whether we for similar photospheric
 pulsation characteristics. 
The photometric data set did not enable us to recover a clear period. 
The long run of observations in a single season with the APT
 covers 131 days (see lower panel Fig.~\ref{phot});
the shallow `secondary minimum', which is a defining feature of the RV\,Tauri
 stars might be present (see lower panel Fig.~\ref{phot}),
 but we lack a long  enough continuous time series to label the light curve as
 due to a genuine RV\,Tauri pulsation.
 The pulsation time-scale seems to be typical for RV\,Tauri stars. The
successive deep minima do not repeat well, however. The full data set has been
analyzed using three period finding methods but no clear conclusions on 
pulsational timescales could be reached. We found two periods of
nearly equal probability, 71.9 and 89.9 days. Additionally, the observations
were plotted against all periods at one day intervals from 67 to 110 days
as well as the double periods near 144 and 180 days and periods of 50.4 and
216 days suggested weakly by the period-finding programs, and no other
plausible periods were found. The substantial scatter, with for
example two exceptionally bright points (JD 2451608 and 9,see upper panel
Fig.~\ref{phot}) somewhat remote
in time from other observations, is observed also in genuine RV\,Tauri stars.
 The observational errors are not negligible compared to a $V$ peak-to-peak
amplitude of only 0.17\,mag. The colors vary such that $B-V$ and $V-I$
 are bluest when
the star is bright; there is no significant phase difference and the
amplitudes are only 0.02 to 0.03\,mag. The $U-B$ plot is dominated by
intrinsic and observational scatter.

 \iras is no genuine high amplitude RV\,Tauri 
star but is variable with a small amplitude on a similar timescale.

\section{Spectral monitoring}

The main purpose of our spectral monitoring is to obtain radial velocity
measurements by cross-correlation of \iras, hence the signal-to-noise
of every spectrum is kept low. As an interesting byproduct, we can study
the variability of large features, like the \halfa-profile.
In the spectra of RV\,Tauri stars line deformation and even line splitting
is observed. They are commonly ascribed to a passage of a shock moving
through the atmosphere
\citep{1989A&A...215..316G,1991A&A...246..490L,1990A&A...237..159G,
1994A&A...292..133F}. A cross-correlation profile is a good tracer of
this motion in the photosphere \citep{2002A&A...386..504M}.
For \iras deformations are sometimes present in the cross-correlation
profiles (Fig.~\ref{vergmask}), but they are small, indicating that the
shock is weak. The Gaussian cross-correlation profiles obtained with a proper mask
for \iras result in precise radial velocity measurements (Fig.~\ref{vergmask}) and
symmetric profiles.

\subsection{Radial velocities}

\begin{figure}
\resizebox{\hsize}{!}{\includegraphics{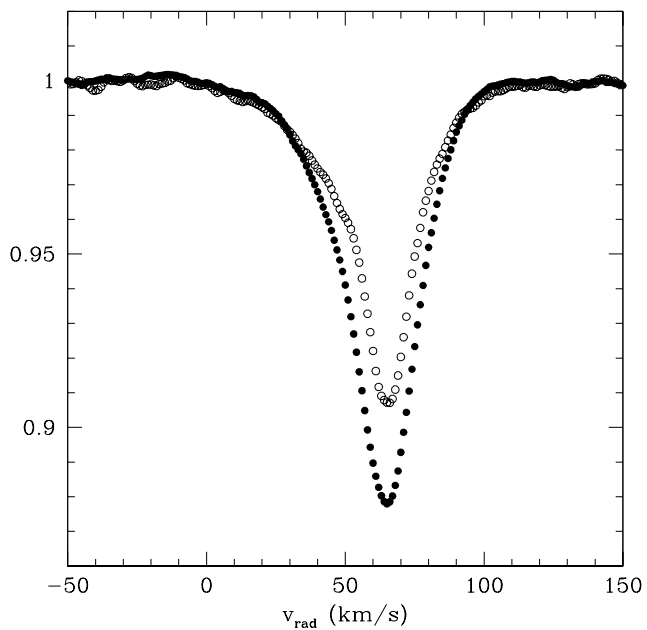}}
\caption{\label{vergmask}Comparison between the cross-correlation profiles
obtained with a standard template (open circles) and a specific template for
 \iras (full circles).}
\end{figure}

The online reduction software for CORALIE provides cross-correlation
profiles computed with a standard template of spectral type F.
As we have a high signal-to-noise FEROS spectrum, we decided to use
this spectrum to construct a proper template tuned for \iras to increase
the efficiency of the cross-correlation technique.
The template spectrum consists of box-shaped emission lines with equal
amplitudes. These lines were taken from the standard template but only the
lines, which are present in the FEROS spectrum and for which we could
determine the FWHM, were selected for the \iras template. We used this
FWHM of the Gaussian fit as the width of the lines.
In this way we reduced the original 3074 line mask to a mask of only 953 lines.
In Fig.~\ref{vergmask} we compare the cross-correlation
profile, computed with the standard template, with the
one computed with the \iras template.
Qualitatively the \iras profiles are more symmetric and the noise in the
profiles is lower, resulting in nicer profiles.
For both templates we fitted the cross-correlation profiles with Gaussians.
The central velocities of the \iras template are shifted with $-$0.17 \kms and a
standard deviation of 0.24 \kms with respect to the velocities of the standard
template.
For the \iras profiles we obtained a mean FWHM of 9.31 \kms and a mean depth of
 0.11, for the standard profiles a mean FWHM of 7.01 \kms and a depth of 0.08.
The \iras template results, thus, in broader and deeper profiles than the
 standard template.

\begin{figure}
\resizebox{\hsize}{!}{\rotatebox{-90}{\includegraphics{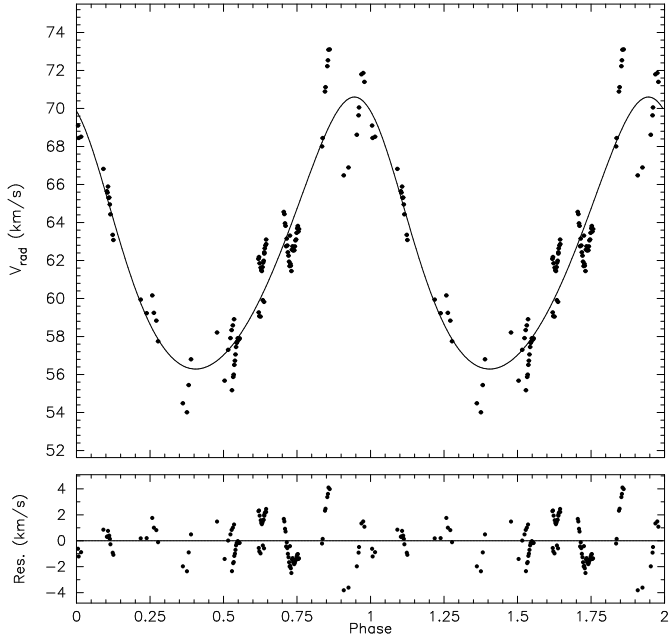}}}
\caption{\label{iras08544rv}
Upper panel : The heliocentric radial velocity data folded on the 499 day
 period. 
The full line is the radial velocity fit discussed in the text. 
Phase 0 corresponds to periastron passage.
Lower panel : the residuals for the radial velocities for the 499 day period 
fit.}
\end{figure}

We have accumulated 107 radial velocities, for which an orbital period
 of 499 days was found. Fig.~\ref{iras08544rv} (upper panel) presents
 the radial velocities folded on the orbital period and the best least-square fit.
 The orbital elements are listed in
Table~\ref{orbelements}. Our dataset covers 2.9 orbital cycles and the
standard deviation of the O-C values for the whole data set is 1.06 \kms.
For the eccentricity we obtained a value of 0.14.
Applying the criterion of \citet{1971AJ.....76..544L}, we find the eccentricity
to be significant (p=6.9 10$^{-3}$ $\%$). Assuming an average inclination of
 60 $^{\circ}$ and a mass of 0.6 M$_{\odot}$
for \iras, we find for the unseen companion a mass of 0.3 M$_{\odot}$.
 The companion could be a late-type main sequence star or a low-mass
(helium) WD. A compact companion makes the evolutionary status of this object
(with the non-zero eccentricity) even more puzzling.

\begin{table}
\caption{The orbital elements of IRAS\,08544. }
\label{orbelements}
\begin{center}
\begin{tabular}{|l|ll|} \hline
 & \multicolumn{2}{|c|}{\iras}  \\
 & & $\sigma$   \\ \hline
\rule[-0mm]{0mm}{3mm}Period (days)          & 499 & 3  \\
a $\sin i$ (AU)        & 0.32 &    \\
F(M) (M$_{\odot}$)     & 0.02 &    \\
K (\kms)               & 7.2 & 0.2 \\
e                      & 0.14 & 0.02 \\
$\omega$ ($^{\circ}$)  & 26  & 10   \\
T$_{0}$ (periastron) (JD245+) & 1964 & 14   \\
$\gamma$ (\kms)        & 62.5 & 0.1 \\
\rule[-3mm]{0mm}{3mm}$\sigma$$_{O-C}$       & 1.06 &  \\  
N                      & 107 & \\ \hline
\end{tabular}
\end{center}
\end{table}

\begin{figure}
\caption{\label{pdm}The $\theta$-statistics of the PDM-analysis for the
 residuals of the radial velocities for a period of 499 days (see text).
 5 bins and 4 covers were used in this analysis.}
\resizebox{\hsize}{!}{{\includegraphics{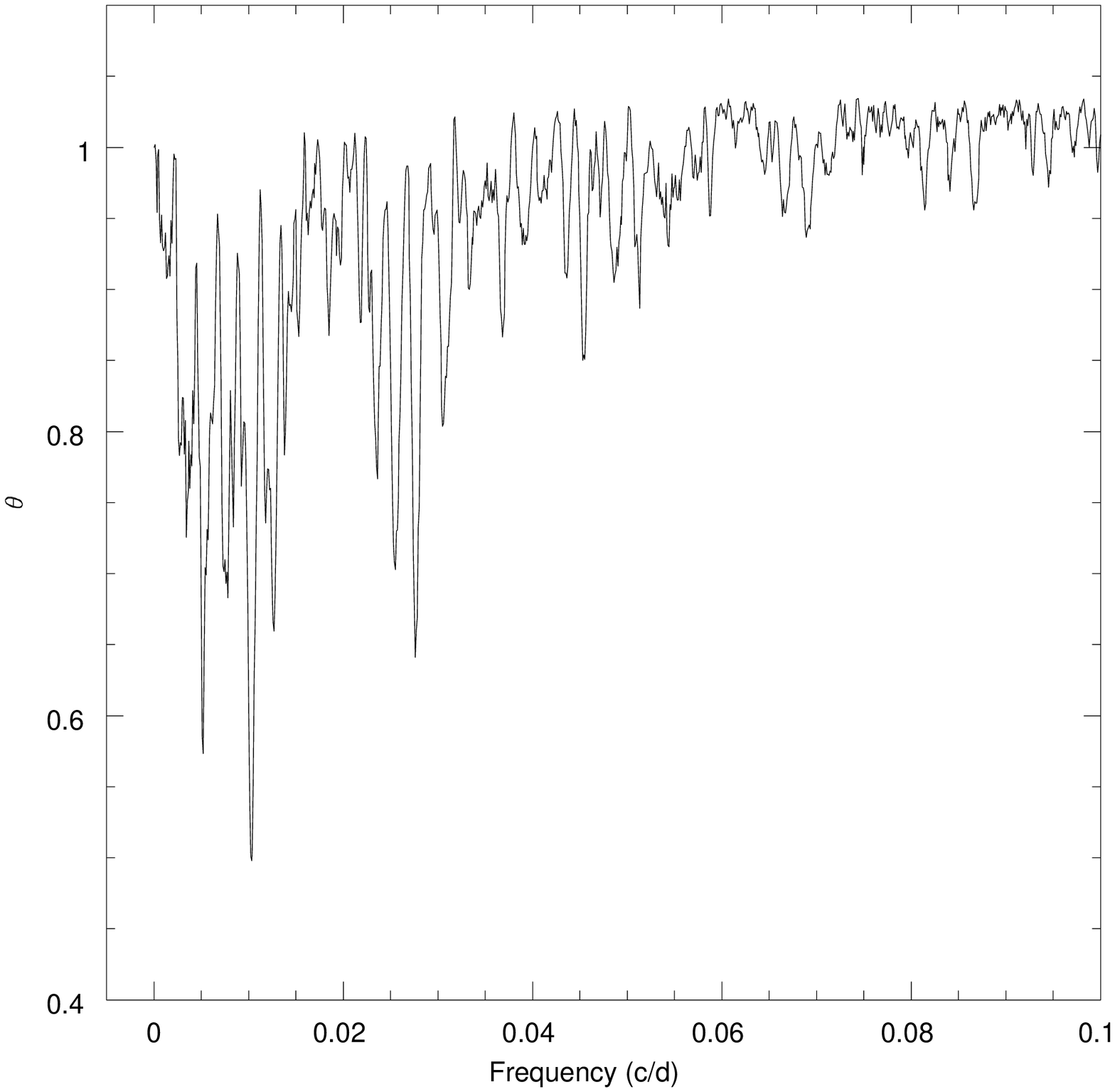}}}
\end{figure}

\begin{figure}
\resizebox{\hsize}{!}{{\includegraphics{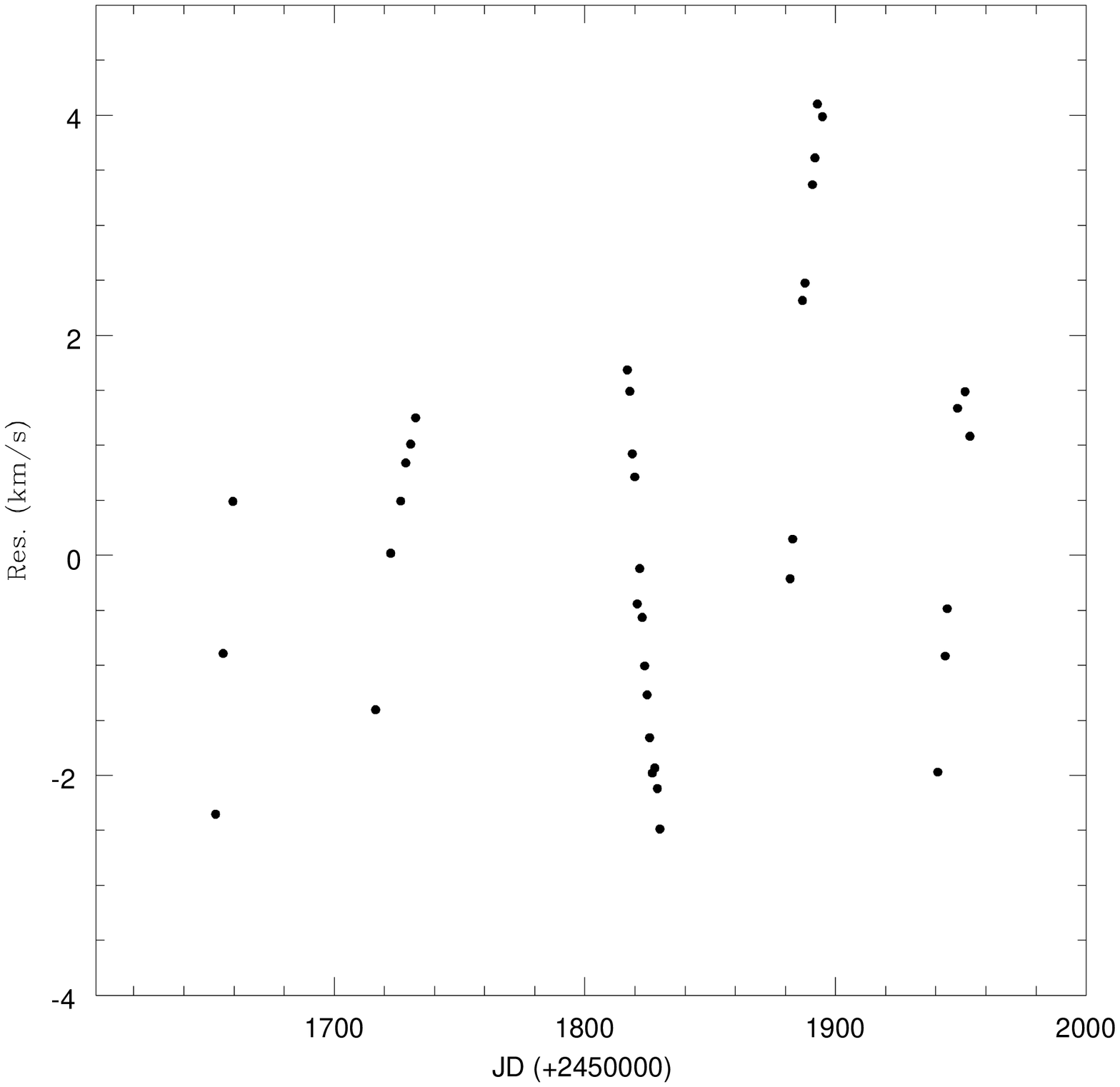}}}
\caption{\label{iras08544res}
The residuals from JD\,2451605 to 2452000, showing a clear structure on a 
timescale shorter than the orbital period. The time-scale of this low amplitude
scatter is around 90 days which we interprete as due to the pulsational amplitude
in the radial velocity data.}
\end{figure}

We have analyzed the residuals (Fig.~\ref{iras08544rv} : lower panel and
 Fig.~\ref{iras08544res})  for periodicities.
 The $\theta$-statistics of the PDM-method \citep{1978ApJ...224..953S} 
 are shown in Fig.~\ref{pdm}.
We suspect variability at a timescale of 100 and/or 30 days, but we
cannot prove yet that this period is stable on a longer time scale.
We interprete this variability  as due to pulsations.

\subsection{\halfa-profile}

\begin{figure}
\resizebox{\hsize}{!}{\rotatebox{-90}{\includegraphics{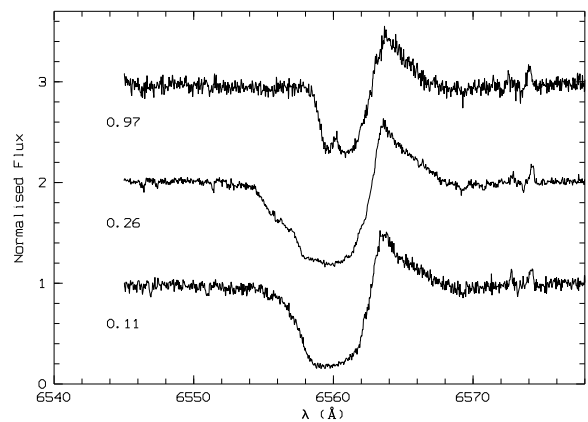}}}
\caption{\label{varhalfa}
The variable P-Cygni profile of \halfa. The orbital phase is indicated for
each spectrum.}
\end{figure}

\begin{figure}
\resizebox{\hsize}{!}{\includegraphics{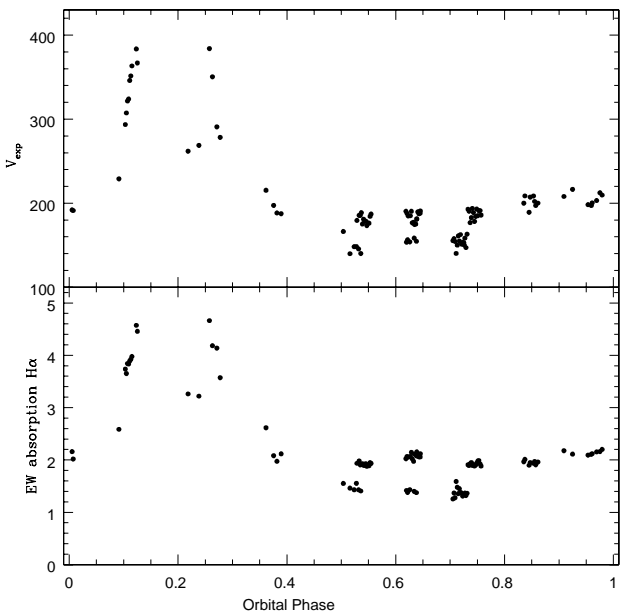}}
\caption{\label{vexpabs}The expansion velocity (upper panel) and the
EW of the absorption component of the P-Cygni profile of \halfa\,(lower panel)
 versus the orbital phase.}
\end{figure}

In all our CORALIE spectra  \halfa\,shows a strong P-Cygni profile of a fast
 stellar wind, which is highly unusual for an F-type object.
The P-Cygni profile varies with time (see Fig.~\ref{varhalfa}) and
indicates that the mass outflow is variable or non-spherically symmetric.
As a detailed model of the outflow is needed to derive accurately parameters
such as the expansion velocity and the mass loss rate, we will only
determine the order of the expansion velocity and describe the observational
characteristics of the profile. We took as the expansion velocity, the
velocity of the intersection of the continuum level and
the tangent in the bending point in the absorption component of the P-Cygni
profile, and found that it varies between 100 and 400 \kms!
The EW of the absorption component (integrated between the wavelength of
the expansion velocity and that of the intersection of the \halfa-line and
the continuum level) varies between 1 and 5 \AA. Fig.~\ref{vexpabs} shows that
both the expansion velocity and the EW of the absorption component of \halfa\
 are correlated with the orbital phase and reach
 a maximum  between phase 0.13 and 0.26. This is around superior
conjunction at phase 0.16 : i.e. when the primary is at its farthest
position with respect to us and the secondary at its nearest.
Thus assuming the F-type star is the \halfa\,emitter in the binary system
, the mass outflow is not spherically symmetric
but seems to reach a maximum in the direction of the companion.
The intensity of the emission peak varies from 1.3 to 1.8 times the continuum
level. As the \halfa-emission intensity is very sensitive to the mass loss rate
\citep{1982ApJ...263..226K}, the mass loss rate of \iras seems to be
variable during orbital motion. Finally, the variation of the EW of the
emission component is smaller than that of the absorpion component and
lies in the range of 0-2 \AA.
In some spectra an extra small blue shifted emission or a slight recovery
 between two absorption components is observed 
(see upper spectrum in Fig.~\ref{varhalfa}).
Just before publication two additional observations were inserted.
 They correspond to orbital phases 0.22 and 0.24 and do not show the high
 expansion velocity and high absorption EW expected from the preceding cycles.
 However, their expansion velocity 
is still higher than those between phases 0.35 and 1.0. This could indicate that
 there is a orbital cycle to cycle  variation in the strength and
 velocity of the \halfa-absorption.


 For the moment neither the origin nor the location in the system of the 
fast  \halfa-emitting gas is understood.

\begin{figure}
\caption{\label{iras08544co} CO data : upper panel : low resolution CO(2-1)
 spectrum taken on 2/10/99, second panel : low resolution CO(2-1) spectrum
 taken on 28/07/00, third panel : total averaged of 5 spectra of the high
 resolution CO(2-1) line , bottom panel : total averaged spectrum of the
 low resolution CO(1-0) line. All velocities are heliocentric.}
\begin{center}
\resizebox{0.4\textwidth}{!}{\rotatebox{270}{\includegraphics{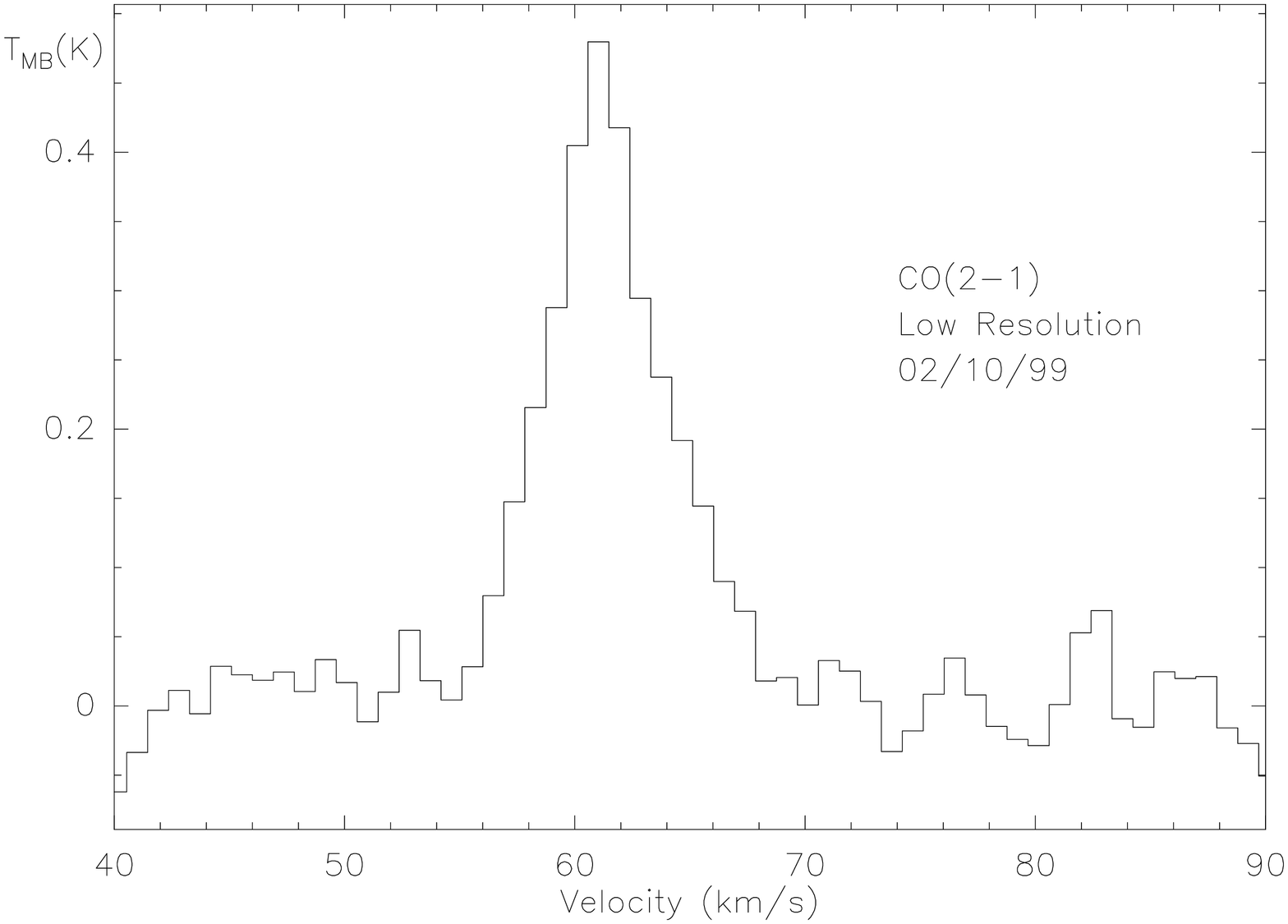}}}
\resizebox{0.4\textwidth}{!}{\rotatebox{270}{\includegraphics{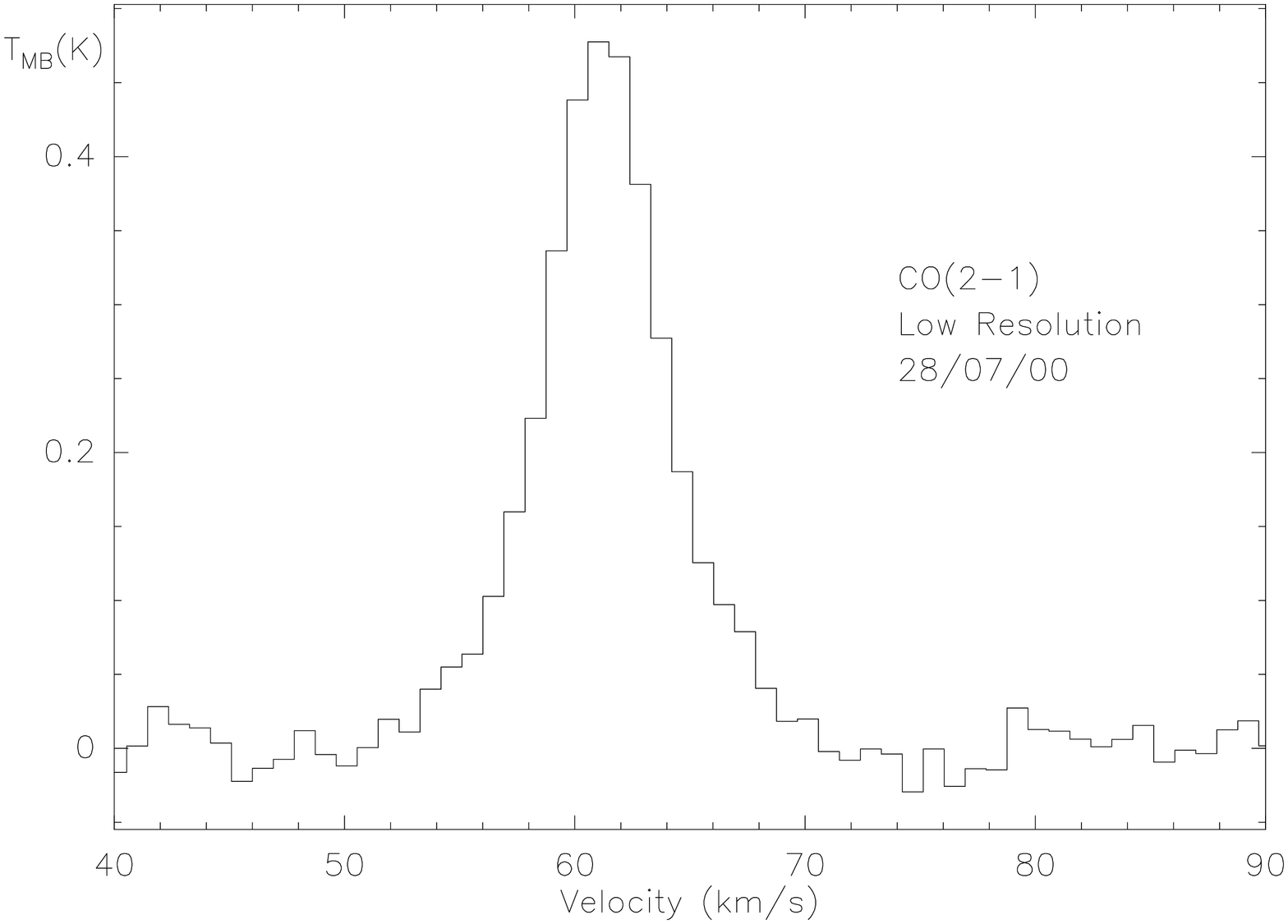}}}
\resizebox{0.4\textwidth}{!}{\rotatebox{270}{\includegraphics{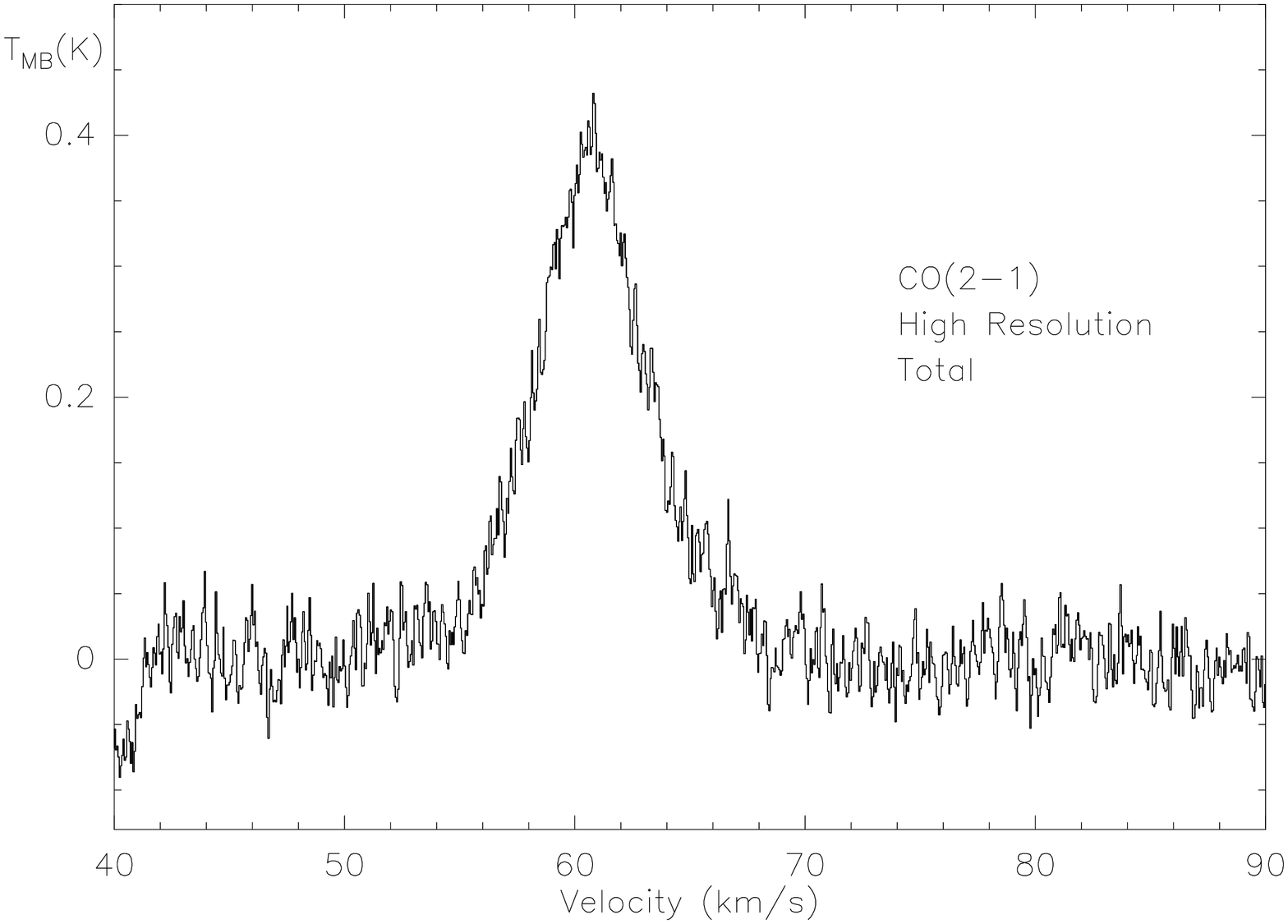}}}
\resizebox{0.4\textwidth}{!}{\rotatebox{270}{\includegraphics{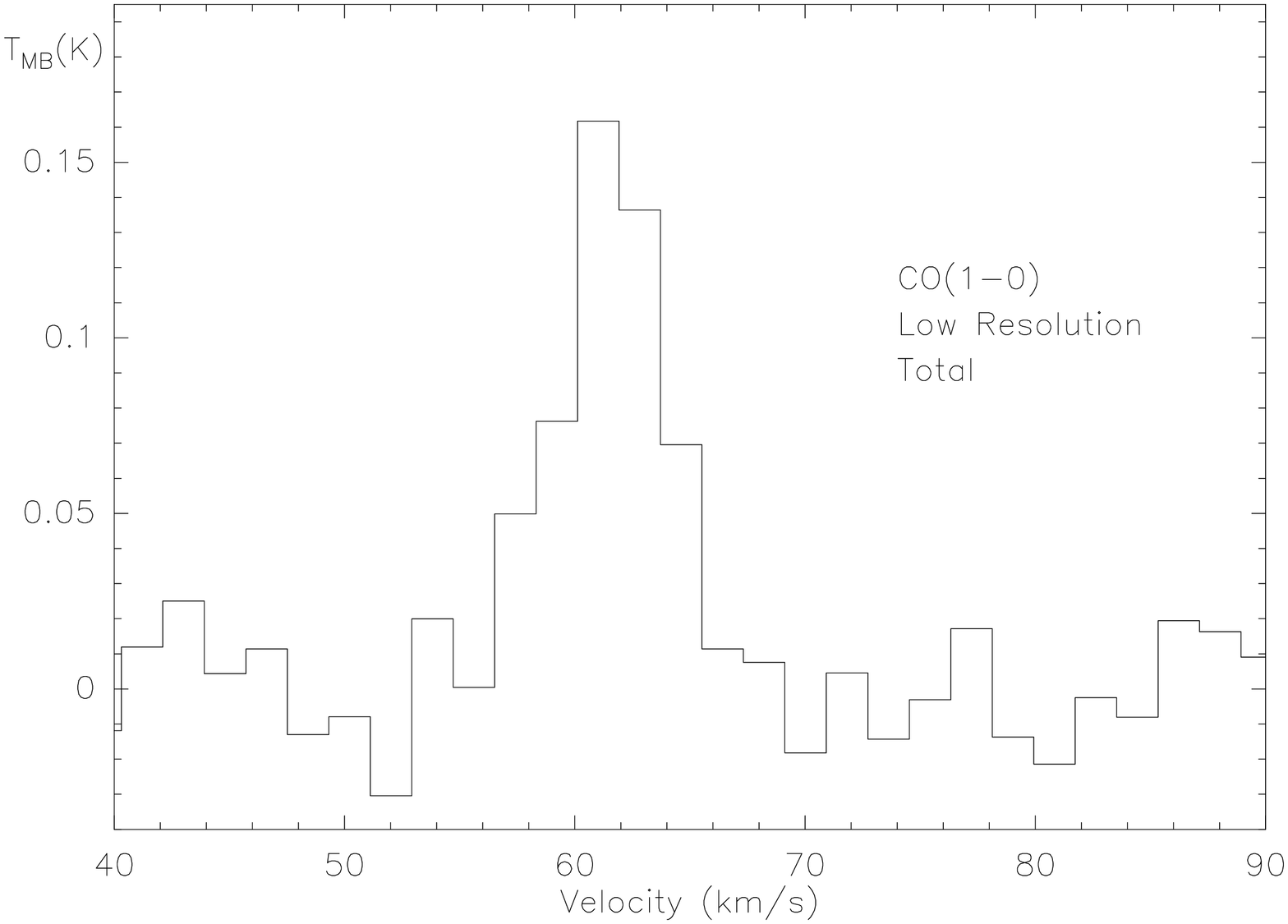}}}
\end{center}
\end{figure}

\begin{figure}
\caption{\label{comap}The map of the environment of \iras in low resolution
 CO(2-1).}
\resizebox{\hsize}{!}{\rotatebox{270}{\includegraphics{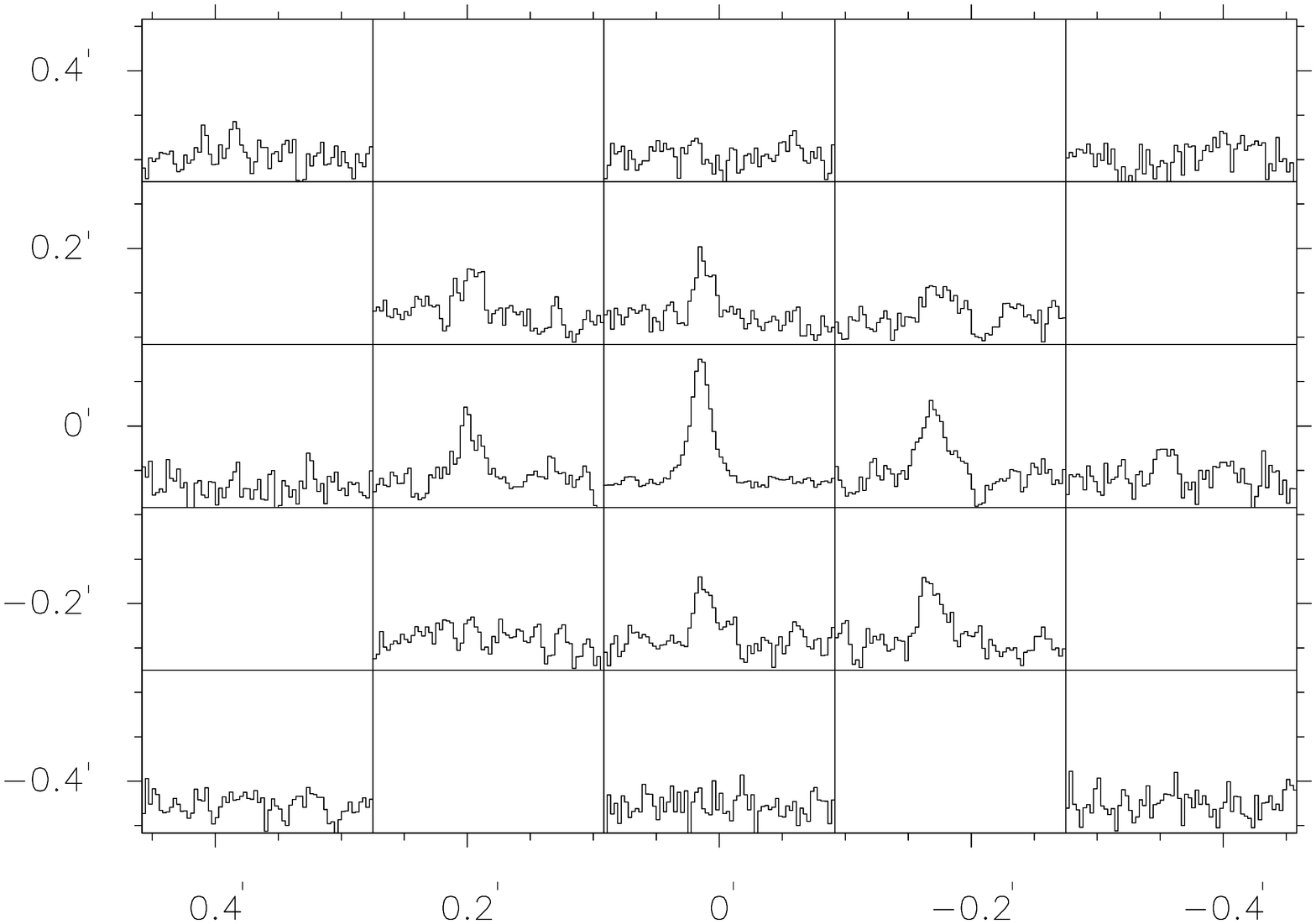}}}
\end{figure}

\section{CO mm-wave line emission}

\subsection{CO (J=1-0) and (J=2-1) emission lines}

\begin{table}
\caption{\label{tabco} The central heliocentric velocity (\kms) ,the FWHM (\kms) and the
integrated intensity (K \kms), obtained by Gaussian fitting, for the different
 lines. The second quoted line is the central
 position of the map shown in Fig.~\ref{comap}.}
\begin{center}
\begin{tabular}{|cc|lll|} \hline
line &   & V$_{c} $ & FWHM  & I$_{CO}$   \\
\hline

CO (2-1) LRS & 2/10/99 & 61.3 & 6.3 & 2.8 \\
CO (2-1) LRS & 28/07/00 & 61.3 & 6.4 & 3.1 \\
CO (2-1) HRS & total & 60.7 & 5.9 & 2.3 \\
CO (1-0) LRS & total & 61.7 & 5.4 & 0.9 \\
\hline
\end{tabular}
 \end{center}
\end{table}

The CO data for \iras are shown in Fig.~\ref{iras08544co}.
For all spectra a baseline is subtracted. The random mean scatter in the
spectra outside the line is of the order of 0.02 K.
In Table~\ref{tabco} the central heliocentric velocities, the FWHM and the
integrated intensities, obtained by Gaussian fitting, for the different lines
 are presented.

The central velocity of the low resolution CO (2-1) line taken on 2/10/99
 (Fig.~\ref{iras08544co}, upper panel), which
corresponds to an orbital phase of 0.94, is  61.3 \kms. However, 
our CORALIE data (see Fig.~\ref{iras08544rv}) point to a  photospheric
 velocity of 70.6 \kms at this orbital phase.
The variation due to pulsation can not explain the difference
between the orbital velocity and the velocity derived from the CO line.
This means that the CO emission is not following the orbital motion of \iras.

Comparing the two low resolution spectra of the CO (2-1) line, observed at
2/10/99 and 28/07/00 (Fig.~\ref{iras08544co}, first and second panel
 respectively), the latter being the spectrum of the
central position of our map, we state that both have the same central
velocities. This is valid for all our profiles.
 We can thus conclude that the central velocity of the CO lines is constant.
 It traces the motion of the system and not that of the primary star.
However, a small difference is found between the central velocity (60.7 \kms,
taken from the averaged HRS CO (2-1) line) and the system velocity of
 62.5 \kms derived from the CORALIE data.


The CO emission of \iras is weak compared to its 60 $\mu$m flux.
For AGB-stars a correlation is observed between the integrated intensity of
the CO (1-0) line and the 60 $\mu$m flux \citep{1992A&AS...93..121N}.
 Also single post-AGB stars like the 21 $\mu$m objects follow this
 correlation \citep{1999A&A...343..202V}. Using
this correlation for oxygen rich stars, the observed  60 $\mu$m flux of 56
Jansky for \iras corresponds to an integrated intensity of the CO (1-0) line
 of 5.5 K \kms.
 This is a factor 6 higher than the
 observed CO (1-0) integrated intensity (I$_{CO}$ =0.92 K \kms).
With respect to the 60 $\mu$m flux coming from the dust, the circumstellar
environment of \iras is deficient in CO, which is common in RV\,Tauri objects
\citep{1991A&A...245..499A,1999A&A...343..202V}.

The triangular line shape of the CO lines is peculiar and differs from
the CO line shape expected for a mass-losing (Post)-AGB star. For 
a spatially unresolved mass-losing (Post)-AGB star the CO line is rectangular
for a low optical depth and parabolic for a high optical depth 
\citep{1993ApJS...87..267O}. 
For 89\,Her, AC\,Her, HD\,44179 (the central star of the Red Rectangle),
 BM\,Gem and EU\,And a similar line profile as for \iras is observed
 \citep{1991A&A...246..153L,1991A&A...245..499A,1988A&A...206L..17B,1995ApJ...453..721J,1999ApJ...521..302J}. All these objects show a weak and small
CO emission component with a FWHM in the range of 0.5-5 \kms and for the latter 
four \citet{1999ApJ...521..302J}
argue that these emission components are the signatures of long-lived reservoirs
of gravitationally bound gas. 
With a FWHM of 6 \kms, \iras falls slightly out of this range,
 but a higher inclination
 angle of the orbital plane can increase the FWHM of the CO emission line.
  
We conclude that the CO line-profile of \iras is also indicative for the 
presence of gravitationally bound gas in a circumbinary disc.
 Note that due to the misclassification of the LRS IRAS spectrum,
 this object was observed and detected also in the CO survey of 
infrared carbon stars by \citet{2002A&A...390..501G}. 
 We checked the profile of \citet{2002A&A...390..501G}, the velocity is the
 same but  they used the half-width-zero-intensity to compute
 the outflow velocity for which they derive 8.2 \kms.
We checked their line-profile which was kindly given to us but since we
obtained a high S/N high-resolution spectrum, we used our detection
 in what follows.

To obtain a rough lower bound for the mass contained in the CO envelope
we follow the train of thought of \citet{1999ApJ...521..302J},
 \citet{1997ApJ...485..341J} and \citet{1998ApJ...500..466K}.
 As an approximation we assume that the gas is in circular orbits around
the binary system for which we adopt a mass of 0.9 M$_{\odot}$ (0.6 M$_{\odot}$
for the Post-AGB star +  0.3 M$_{\odot}$ for the companion).
We estimate the inner radius, $R_{in}$ of the disc from the CO line FWHM.
If $V_{orb}$ denotes the orbital velocity of the gas,

\begin{equation}
R_{in} = GM/V_{orb}^{2}
\end{equation}

 With $V_{orb} = 3$ \kms\  from our CO (2-1) high resolution line,
$R_{in}= 90 AU$. The luminosity of \iras lies in the range
of 3000-7000 L$_{\odot}$. We adopt here a value of 5000 L$_{\odot}$.
At a distance of 90 AU  from the star with a luminosity of
 5000 L$_{\odot}$ a blackbody grain has a predicted temperature of 250 K.

If the gas is optically thin we can crudely estimate the column density of
 CO molecules in the beam,
 $\overline{N(CO)}$

\begin{center}
\begin{equation}
\overline{N(CO)}=\frac{3k^{2}T_{ex}}{4\pi^{3}\mu^{2}h\nu^{2}}\int{T_{mb}dv}
\end{equation}
\end{center}

where $\mu$ is the dipole moment of the CO molecule (0.1098 D; \citealt{debye})
 and $\nu$ is
 the frequency of the (2-1) transition (230.538 GHz). In this equation
several assumptions have been made : (1) the microwave background is ignored,
 (2) the partition function scales as $kT_{ex}$ and (3) the Planck function and
 the correction term for stimulated emission can be simplified by linear
 expansion of $exp(\pm h\nu/kT_{ex})$.
 We find
 $\overline{N(CO)} \simeq 6 \times 10^{15} cm^{-2}$.

The total mass of CO molecules ($M_{CO}$) in the beam is the product of
 $\overline{N(CO)}$, the mass of a CO molecule ($m_{CO}$)
 and the projected area of the beam on the sky $A_{beam}$.
 For a beam assumed to be Gaussian with half-power diameter $\theta_{HPBW}$,

\begin{equation}
A_{beam}= \pi\theta^{2}_{HPBW}D^{2}_{\ast}/(4ln2)
\end{equation}

with $D_{\ast}$ the distance to the star. If we adopt a distance of 1 kpc,
we find M$_{CO} = 3.6 \times 10^{25} kg $. With the assumption of $ [CO/H]=10^{-4}$,
the total mass in the CO envelope is 0.2 M$_{\odot}$. If the gas is optically
 thick, this value is a lower bound.

\subsection{CO map}

We mapped the environment of \iras in low resolution CO(2-1).
The CO map consists of 14 positions with a spacing of 11'' and is presented in
 Fig.~\ref{comap}. The clear maximum observed in the central position of the 
map, which  is also presented in the second panel of
Fig.~\ref{iras08544co}, indicates that the CO line is not interstellar.
 We find that an asymmetry is present in the map: while
positions (-1,1) and (1,-1) show a clear CO emission, the intensity of the
CO emission is lower in position (1,1) and CO emission is absent  in position
(-1,-1). This could mean that the CO emission is anisotropic and extended.
 However, this asymmetry can also be caused by pointing errors in 
the different positions of the map, which can also explain the signal
present in the position (2,0).


\section{Chemical composition}

\begin{figure}
\resizebox{\hsize}{!}{\includegraphics{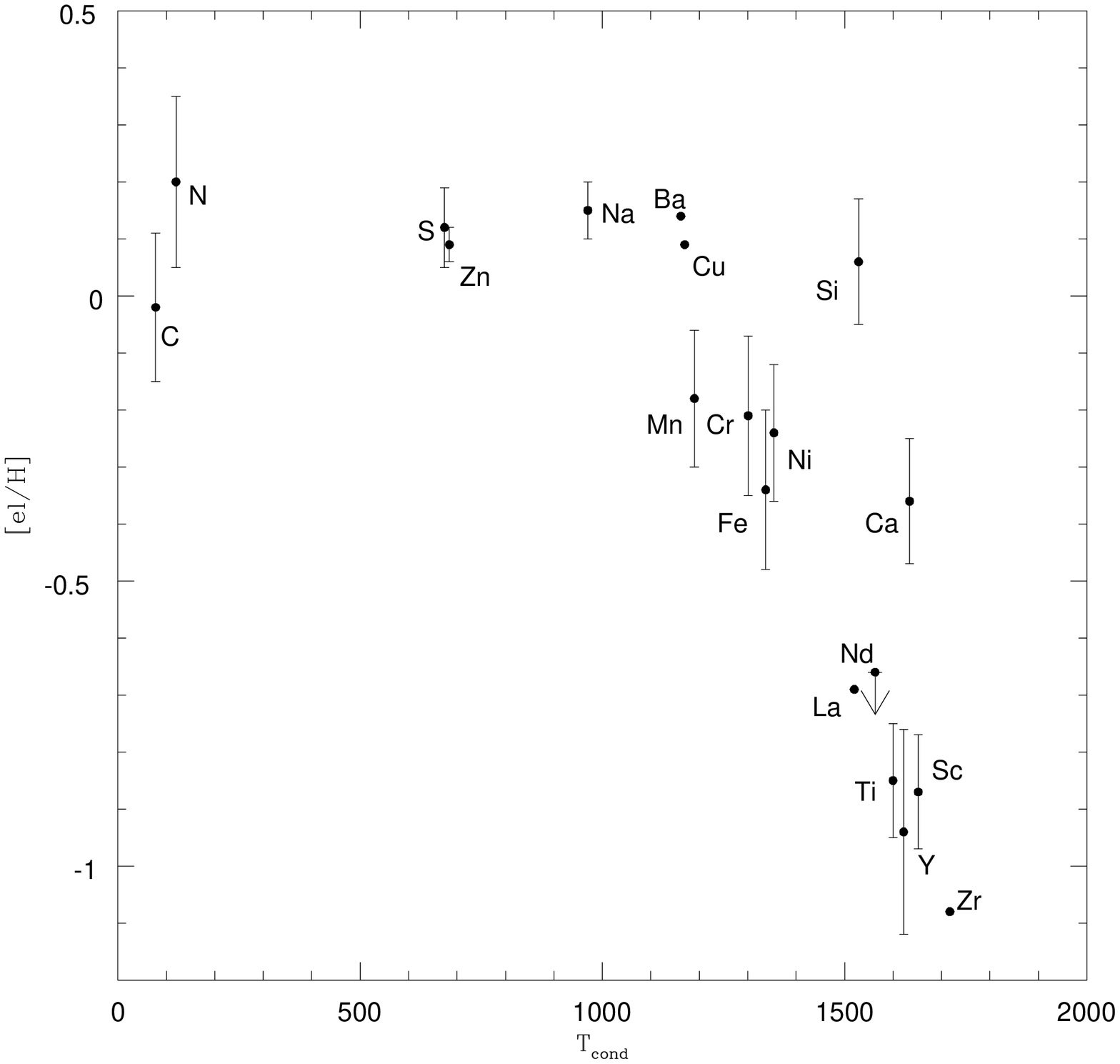}}
\caption{\label{irascondens} The composition of \iras against the dust
condensation temperature of the chemical element.}
\end{figure}

\subsection{Atmospheric Parameters}

The abundances are calculated on the basis of a LTE model atmospheres of
 \citet{1993KurCD..13.....K} and the LTE program MOOG of Sneden.
A model atmosphere is determined by its effective temperature ($T_{eff}$),
 the gravity ($\log g$) and the metalicity ([Fe/H]).
We determined these parameters on the basis of spectroscopic criteria :
the effective temperature by forcing the abundances of Fe I-lines to be
 independent of the excitation potentials
and the gravity by forcing ionization balance between the Fe I and Fe II lines.
 Moreover, the microturbulent velocity was estimated by forcing the abundance
 of the FeI-lines to be independent of the reduced equivalent
width. We then used the model atmospheres with the correct overall metalicity
 as given by the Fe abundance.
The oscillator strengths of the lines were taken from the critically compiled
database continuously updated in our institute. For more details we refer to
\citet{2000A&A...354..135V}.

We obtained $T_{eff}$ = 7250 K, $\log g$ = 1.5 and  [Fe/H] = $-$0.5 for the
model parameters. This temperature is in agreement with
the spectral type F3 determined from low resolution spectra and confirms the
position of \iras on the high temperature side of the
instability strip.

\subsection{Results}

\begin{figure}
\resizebox{0.45\textwidth}{!}{\rotatebox{-90}{\includegraphics{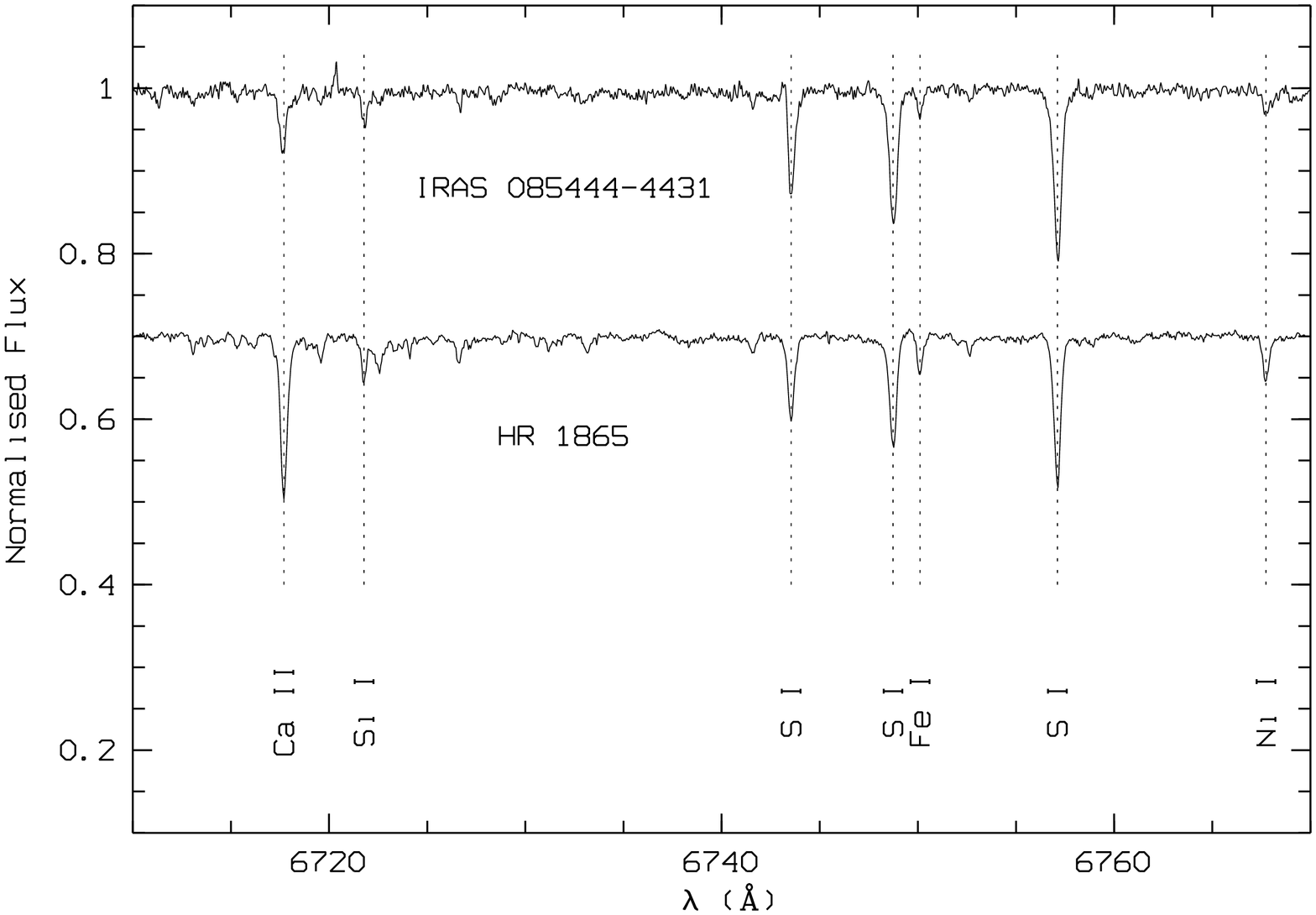}}}
\caption{\label{spec6740}Sample spectra of \iras and HR\,1865. The spectra
are velocity corrected. HR\,1865 is a massive supergiant with the same
  atmospheric parameters. All lines in HR\,1865 are stronger than in \iras
 except for the lines of S.}
\end{figure}

\begin{figure}
\resizebox{0.45\textwidth}{!}{\rotatebox{-90}{\includegraphics{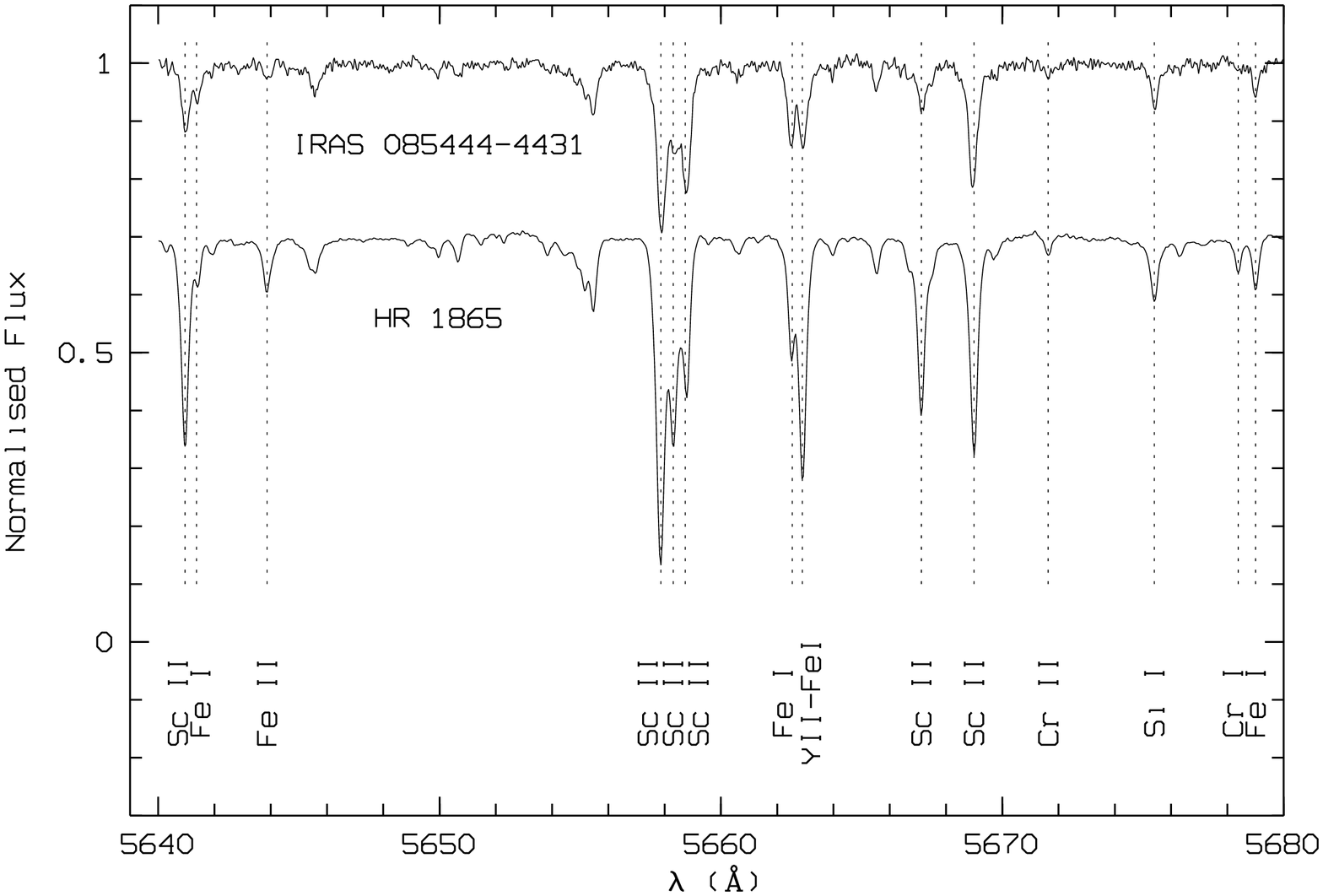}}}
\caption{\label{spec5660}Spectrum of \iras around 5660 \AA. All lines in
HR\,1865 are stronger than in \iras, but the difference in Sc-lines is largest.}
\end{figure}

The abundance results obtained for \iras are listed in Table~\ref{chem}.
In Fig.~\ref{irascondens} the abundance of the elements are plotted versus
 their dust condensation temperature.
The spectrum has a high S/N of 150 (measured around 5800 \AA) and we used
a total of 264 lines, restricting ourselves to small lines (EW $\leq$ 150 m\AA).
 These lines have, like all the other lines
in the spectrum, symmetric profiles (see Fig.~\ref{spec6740} and~\ref{spec5660}).
The good consistency between the abundances derived from the different
ionization levels for the elements Cr, Mn, Ni confirms the reliable model 
atmosphere parameters.

\begin{table}
\caption{\label{chem}Chemical analysis of \iras. For the solar iron abundance
we used the iron meteoric abundance of 7.51. For the solar C,N and O
abundances we adopted resp. 8.57, 7.99 and 8.86
(C: \citealt{1993ApJ...412..431B}, N: \citealt{1991A&AS...88..505H},
O: \citealt{1991ApJ...375..818B}). For the solar
 Mg and Si abundances we adopt 7.54 \citep{2001sgc..conf...23H};
 the other solar abundances are taken from \citet{1998SSRv...85..161G}.
 The dust condensation
temperatures are from \citet{Lodders} and references therein.
 They are computed using a solar
abundance mix at a pressure of 10$^{-4}$ atm. They do not ascribe a
 condensation temperature to O as
it is the most abundant element in rock and therefore a separate condensation
 temperature is meaningless. For all ions, the number of lines, the mean 
equivalent width, the absolute abundance and the $\sigma$ of the line-to-line
scatter are given.}
\vspace{0.5ex}
\begin{center}
\begin{tabular}{lrr|rlrl}\hline
\multicolumn{7}{c}{\rule[-0mm]{0mm}{5mm} {\iras}}\\
\multicolumn{7}{c}{ {\em $T_{eff}$=7250 K}}\\
\multicolumn{7}{c}{ {\em $\log g$=1.5}}\\
\multicolumn{7}{c}{{\em $\xi_t=4.50$ \kms}}\\
\multicolumn{7}{c}{\rule[-3mm]{0mm}{3mm} {\em [Fe/H]=$-$0.5}}\\
\hline
\rule[-3mm]{0mm}{8mm}ion & N & $\overline{W_{\lambda}}$ & $\epsilon$ &
$\sigma$  & [el/H] & T$_{cond}$\\
\hline
\rule[-0mm]{0mm}{3mm}C~I  & 21 & 68 & 8.55 & 0.13 & $-$0.02 &78 \\
N~I   & 3 & 81 & 8.19& 0.15 & 0.20    &120\\
O~I   & 4 & 49 & 8.67& 0.05 & $-$0.19    &  \\
Na~I  & 6 & 42 & 6.48& 0.05 & 0.15    &970\\
Si~I  & 9 & 26 & 7.60& 0.11 & 0.06    &1529\\
S~I   & 6 & 56 & 7.45& 0.07 & 0.12    &674\\

Ca~I  &13 & 69 & 6.00& 0.11 & $-$0.36 &1634\\
Sc~II & 5 & 85 & 2.30& 0.10 & $-$0.87 &1652\\
Ti~II &16 & 75 & 4.17& 0.10 & $-$0.85 &1600\\
Cr~I  &11 & 34 & 5.46& 0.15 & $-$0.21 &1301\\
Cr~II &16 & 91 & 5.46& 0.13 & $-$0.21 &\\
Mn~I  & 5 & 50 & 5.22& 0.12 & $-$0.17 & 1190\\
Mn~II & 1 & 41 & 5.20&      & $-$0.19 & 1190\\
Fe~I  & 83& 55 & 7.19& 0.13 & $-$0.32 &1337\\
Fe~II & 39& 66 & 7.16& 0.15 & $-$0.35 &\\
Ni~I  &16 & 29 & 6.03& 0.12 & $-$0.22 &1354\\
Ni~II & 1 & 27 & 6.00&      & $-$0.25 &1354\\
Cu~I  & 1 & 19 & 4.30&      & 0.09    &1170\\
Zn~I  & 2 & 81 & 4.69& 0.03 & 0.09    &684\\
Y~II  & 3 & 16 & 1.30& 0.18 & $-$0.94 &1622\\
Zr~II & 1 & 6 & 1.52&  & $-$1.08 &1717\\
Ba~II & 1 &131 & 2.27&      & 0.14    &1162\\
La~II & 1 & 13 & 0.48&      & $-$0.69 &1520\\
\rule[-3mm]{0mm}{3mm}Nd~II & & &    &$<0.84$& $<-$0.66&1563\\\hline
\end{tabular}
\end{center}
\end{table}

With [Fe/H]=$-$0.3 and [Zn/Fe]=$+$0.4 \iras displays a weak signature
for depletion of refractory elements in its photosphere.
In Fig.~\ref{irascondens} we can distinguish two groups of elements.
For elements like C, N, S, Zn, Na, Ba, Cu there is no correlation between
the abundance and the condensation temperature. For elements with a condensation
temperature higher than about 1200 K the abundance goes down from
$-$0.2 for
[Cr,Mn,Ni/H] to $-$1.1 for [Zr/H], as the condensation temperature increases.
One element doesn't really fit in this picture which is Si.
\citet{Lodders} give a condensation temperature of 1529 K for Si. However
 in the notes they say that most Si condenses into $MgSiO_{3}$ and
 $Mg_{2}SiO_{4}$ which has a condensation temperature of 1340 K. This
lower temperature would improve the position of Si in our
 interpretation of Fig.~\ref{irascondens}.

To verify the depletion pattern observed in Fig.~\ref{irascondens}, we also
calculated the abundances on the basis of two other model atmospheres.
 The first model has the parameters $T_{eff}$ = 7500 K, $\log g$ = 2.0 and
[Fe/H] = $-$0.5 , the second the parameters  $T_{eff}$ = 7000 K, $\log g$ = 1.5
 and  [Fe/H] = $-$0.5. For the first model there is a global upward shift
with a mean of 0.16, for the second model a global downward shift with a mean
of 0.13. However, the typical depletion pattern is conserved for both models.
Moreover, the abundance ratios [S/$\alpha$] ($\alpha$ = Si,
 Ca and Ti) and [Zn/Fe] which trace the
depletion of a photosphere, are almost independent of the model used.
For the first model we obtained [S/$\alpha$]=0.45 and [Zn/Fe]=0.42,
 for the second [S/$\alpha$]=0.54 and [Zn/Fe]=0.35
 (compared to [S/$\alpha$]=0.50 and [Zn/Fe]=0.43
 for our preferred model). So we can conclude that the depletion is small but
real.

The small N-overabundance ([N/H]=$+$0.2) might be the consequence of
 a first dredge-up. As the C-abundance is solar and
the s-process elements follow the depletion pattern, there is no
evidence for dredge-up of material processed by nucleosynthesis
on the AGB. 

 For the s-process element Nd we could only derive an upper limit since
even the strongest expected lines were not present in the spectra.
The upper limit is therefore derived assuming a fixed EW of 5 m\AA\ for the
 strongest line in a
clear part of the spectrum. We used the line lists of the Vienna Atomic Line
 Database
(Vald2, http://www.astro.univie.ac.at/vald/, \citealt{1999A&AS..138..119K}) to look for
these strongest optical lines.

We also noted a P-Cygni profile for two FeI lines, at 5576.089 \AA \, and at
6400.001 \AA.
We also detected 9 emission lines in the spectrum (7 of FeI, 1 of CaI and
1 of TiI). All these lines have low excitation potentials.

To summarize, \iras is a star where the depletion process has slightly altered
the photospheric abundances ([Zn/Fe] = +0.4) and for which there is no evidence for
typical 3rd dredge-up chemical enrichement.

\section{Conclusions and discussion}

The main conclusions obtained from the analysis of our large data-set are : \\

1) \iras is not a carbon star as listed in the literature but an
F-type post-AGB star with IR colours similar to the bulk of the IRAS detected
RV\,Tauri stars; \\
2) The star shows variability with a small amplitude ($\Delta V$=0.17) for 
which the time-scale (between 30-100 days) is not repeating well; \\
2) \iras is a binary system with a period of 499 $\pm$ 3 days and a significant
non-zero eccentricity of 0.14 $\pm$ 0.02; \\
3) the SED shows a broad IR-excess, indicating beside the presence of cold
 circumstellar dust also the presence of a strong hot dust component indicative 
of a circumbinary disc; \\
4) the weak CO emission line and its triangular shape also points
to a circumbinary location with a low gas-to-dust ratio; \\
5) The \halfa-line shows a P-Cygni profile for which both the strength and outflow velocity 
are correlated with the orbital period with maxima at superior conjunction;  \\
6) the photosphere is moderately depleted in refractory elements ([Zn/Fe]=+0.4). \\

\iras was selected on the basis of its IRAS colours which fall into the 
RV\,Tauri box as defined by \citet{1985MNRAS.217..493E}.
\citet{1997ApJS..112..557K} classified IRAS sources on the basis of their IRAS
 low-resolution spectra. They report the detection of circumstellar 11\,\mic\
SiC dust emission for \iras and classify the star as a possible carbon star. 
Our opical data clearly identifies the star as an F-type object and also
the oxygen rich nature of the photosphere derived in our chemical abundance analysis
does not agree with a post-carbon star spectrum.
Examination of the LRS spectrum \citep{1986A&AS...65..607O} shows that the
18\,\mic\ silicate band is present in the form of weak emission, 
while the structure near 10\,\mic\
may be understood as the 9.7\,\mic\ silicate band with central reversal
and low contrast. The spectrum therefore shows self-absorption. An
alternative classification on the original scheme of
\citet{1986A&AS...65..607O} would be 31, indicating silicate absorption.
Silicate self-absorption is explicitly taken into account in the system of
\citet{1990MNRAS.243..336E}, where the appropriate classification would
be 21BA. The equivalent on the new KSPW system \citep{2002ApJS..140..389K}
is 3.SB or 4.SB. This reclassification removes the anomaly.

\citet{1999IAUS..191..453E} determined an RV\,Tauri
 instability strip based on the RV\,Tauri
stars listed in the GCVS. The amplitude of \iras is small for an RV\,Tauri star
 and its spectral type places it outside and on the hot side of this
instability strip. This suggest that the star has almost left the classical 
instability strip towards the phase of the planetary nebulae. Despite its probably
more evolved nature, the infrared colours of \iras are very similar to the 
IRAS detected RV\,Tauri stars. 

Moreover, \iras is a binary with a period of 499 days together with a 
significant eccentricity of 0.14 which is difficult to explain in terms of 
a previous standard binary AGB-evolution.
On the AGB, the orbit was too small to accommodate an M-star with the same 
luminosity and it is not clear how the star avoided spiral-in, 
leading to a circular orbit of short period. 
Post-AGB binaries are not uncommon and other objects with similar orbits
and large eccentricities were observed (e.g. \citealt{1997IAUS..180..313W,vanwinckel2003}). 
Population synthesis codes have shown that systems of post-AGB stars with
 resulting periods in the order of one hundred to a few hundred days can only be
 expected if an efficient tidally enhanced mass-loss is induced. 
This wind increases the mass-loss significantly 
prior to Roche-lobe overflow (RLOF) and decreases the envelope binding energy to
avoid spiral-in to very short orbits.  Following \citet{pols2003} no 
eccentric systems with P $<$ 3000 days are produced; this is at variance with 
earlier reports by \citet{2000MNRAS.316..689K}.  \iras is another binary 
with a rather short
period and eccentric orbit in an apparent post-AGB evolutionary phase for which 
the mere existence is not very well understood 
in binary stellar evolution theory.

It has been known
for some time that a subset of those post-AGB binaries are extremely depleted 
(e.g. \citealt{1995A&A...293L..25V}). The depletion process implies that the star is
coated with a gaslayer, from which the dustgrains were
removed, probably by radiation pressure. Since dust formation is
 thought to occur in the circumstellar environment and
 not in the photosphere, the process involves
re-accretion of circumstellar gas. \citet{1992A&A...259L..39M} have discussed
several processes in which separation of circumstellar dust and gas may operate
and \citet{1992A&A...262L..37W} added that the most favorable circumstances may occur if the
circumstellar dust is trapped in a stable disc. For post-AGB stars
the presence of a circum-system disc implies binarity and the observational
basis of this process was that all strongly depleted
objects known at that time were found to be binaries with strong
evidence for some of them that the dust was indeed in a flattened
geometry \citep{1995A&A...293L..25V}.

The binary prerequisite for the depletion process was disputed, however, after
the recent  finding that many RV\,Tauri stars are also depleted. 
\citet{1963ApJ...137..401P} subdivided
the RV\,Tauri stars into three spectroscopic
classes (RVA,RVB,RVC) : RVA have spectral type G and K and show strong metallic
 lines and have normal CN and CH bands. RVB have spectral type Fp(R), they
are weak-lined but show enhanced CN and CH bands. RVC are also weak lined
 objects with spectral type Fp but with normal CN and CH bands. There is now
general agreement that the abundance anomalies are mainly due to a depletion process
and are not reflecting internal dredge-up processes 
\citep{1994ApJ...437..476G,1998ApJ...509..366G,2000ApJ...531..521G,
1997ApJ...481..452G,1997ApJ...479..427G,1998A&A...336L..17V,2002A&A...386..504M}.
The strength of the abundance anomalies differs 
from star to star. However, all the stars of the RVB class are affected,
 while the RVA class contains stars which are less or not affected at all.
The RVC stars, which have a low initial
 metallicity [Fe/H]$\le$-1.0, do not show any sign of
depletion \citep{2000ApJ...531..521G}. In IRAS\,08544, the depletion altered the photospheric
chemical composition but only mildly so.
This is well established by the significant ratios [Zn/Fe]=$+$0.4,
 [S/$\alpha$]=$+$0.5. As we have evidence for a current outflow,
 no accretion of gas is going on. The fractionation process has thus 
occurred in the past. However, the atmosphere is only slightly changed in its
 composition and since \iras shows a large and
broad excess , this illustrates again that no correlation exists between the
strength of the {\sl current} IR-excess and the severity of the depletion
\citep{2000ApJ...531..521G}.
 An explanation put forward for the moderate  depletion
 of the RVA RV\,Tauri stars is convection  \citep{1997ApJ...481..452G} :
 a depletion process has occurred  but mixing
with deeper layers with normal abundances, has diminished the effects.
Convection could as well  have diluted the anomalies caused by the
 depletion process in \iras, when \iras was cooler and had a deeper convective
envelope.

\citet{1999A&A...343..202V} argued that many 
IRAS selected RV\,Tauri stars are suspected binary systems where the 
circumstellar material resides in a stable dusty disc. 
Such a disc was resolved in a prototypical
RV\,Tauri star AC\,Her \citep{2000ApJ...541..264J} which was found to be a binary
with a period of 1200 days \citep{1998A&A...336L..17V}.
Recently, radial velocity monitoring showed that two strongly depleted
RV\,Tauri stars, RU\,Cen and SX\,Cen, are indeed members of binary systems
\citep{2002A&A...386..504M}.
 Moreover, the orbital period of SX\,Cen is around 600 days, which is similar
to the period of the long term photometric variation in the light curve.
This indicates that the long term photometric variation in SX\,Cen
 is due to variable circumstellar reddening and that the
photometric class of the RV\,Tauri stars has its origin in a difference in
inclination angle : the RVb RV\,Tauri stars are seen near edge on and the
variable circumstellar reddening is due to the circumbinary disc, which is
in the line of sight. The RVa RV\,Tauri stars are seen face on
and no long term photometric variation is present \citep{1999A&A...343..202V}.
The binary system \iras clearly fits in this picture. The
hot dust observed in \iras was not ejected recently but in a previous
phase in which it was stored close to the star.

Another interesting analogue to \iras is
89 Her. It is a semiregular variable (pulsation period = 62 days,
\citealt{1981ApJ...243..576F}) of spectral type F2. It is a binary system 
with an orbital period
of 288 days and an eccentricity of 0.19 \citep{1993A&A...269..242W}.
 The SED shows a broad IR-excess and
the CO emission profile has a very narrow central peak with a width
of less than 1 \kms \citep{1991A&A...246..153L,1991A&A...245..499A}.
Interferometric CO observations of 89 Her reveal the presence of two
distinct shells \citep{1995A&A...303L..21A}. The outer shell lies at a distance
of about $5 \times 10^{16}$ cm from the star and corresponds
 to the mass ejection in the
AGB-phase, which stopped more than 1000 years ago (using an expansion
 velocity of 7 \kms). The inner shell has an outer radius of $ 10^{16}$ cm.
The \halfa-line shows also a P-Cygni profile with an outflow velocity of
150 \kms \citep{1969mlfs.conf...57S}.
The chemical analysis of the photosphere
performed by \citet{1990ApJ...357..188L}
proves, however, that no depletion process has taken place in 89 Her.
\citet{1993A&A...269..242W}  proposed for 89 Her
a model in which dust has been stored
in a circumbinary disc while no dust is formed in the low density outflow. In
the plane of the disc the ejected material is decelerated.
 In the polar regions there is a free outflow. A
similar model could be applicable to \iras. However, our \halfa\ monitoring 
indicates that the maximal outflow
velocity is  towards the companion, not the poles (in which case
a constant outflow velocity would be observed).
Emission lines are observed in the spectrum of 89\,Her, as they are in \iras.
These emission lines could be collisionally excited as the wind of the central
star runs into the circumsystem disc \citep{1993A&A...269..242W}.

On the basis of the triangular shape and the weakness of the CO line
\iras is similar to other similar objects : AC\,Her, HD\,44179 
(the central star of the Red Rectangle), BM\,Gem and EU\,And. 
In these objects (a part of) the CO emission
is coming from an orbiting molecular reservoir \citep{1999ApJ...521..302J}. 
The objects BM\,Gem and HD\,44179 show, beside a small emission component, also
a broad weaker component, probably coming from a bipolar outflow.
 As the CO disk component is extremely small, the two components
are clearly distinguishable.
It is tentatively suggested that the CO profile of \iras also consists
of two components which are difficult to distinguish.
For \iras the CO disc component is wider and is more difficult to
separate from a possible wide component.

Beside a weak and small CO emission component, there are also other observations
for HD\,44179 and the RV\,Tauri star AC\,Her which indicate that
the physical and chemical conditions in a circumbinary disc are different from
those observed in an outflow : ISO-SWS spectra
reveal a high abundance of oxygen-rich crystalline silicates
 \citep{1998Natur.391..868W,1999A&A...350..163M} and a very strong
millimeter continuum flux from large dust grains is present
\citep{1994A&A...285..551V,1995MNRAS.273..906S}.
 IR spectra are needed to verify whether these
features are also present in \iras.
 The photospheres of these three objects are depleted,
 but with variable strength
 : [Fe/H]= $-$0.3 and [Zn/Fe] = $+$0.4 for \iras,  [Fe/H] = $-$1.7  and
[Zn/Fe] = $+$0.7 for AC\,Her
\citep{1998A&A...336L..17V} and [Fe/H]=$-$3.3 and [Zn/Fe]=$+$2.7
  for HD\,44179 \citep{1996A&A...314L..17W}.
The conditions under which the depletion process (whether or not covered by
mixing) is highly efficient, of average efficiency or not at all efficient,
is still not clear and needs further investigation.

We conclude that \iras is an interesting post-AGB binary with many characteristics
of RV\,Tauri stars and which evolved probably just outside the instability strip.
More detailed study, especially at high angular resolution
of \iras and other similar objects will help us
 to clarify the relation between the characteristics and evolution of the circumstellar
environment, the binary nature of the object and the presence of
the depletion process. Moreover, the evolutionary link between the different 
post-AGB binaries needs further investigation.

\begin{acknowledgements}

We are indebted to Greg Roberts for some of the $UBVRI$ photometry with
the 0.5\,m telescope and the Geneva Observatory for the generous awarding
of telescope time on the Euler telescope as well as for the reduction and cross-correlation
 software. We thank Laurent Eyer, Katrien Kolenberg, Bart Vandenbussche, Maarten
 Reyniers, Katrien Uytterhoeven, Tinne Reyniers, Roeland Van Malderen, Caroline
Van Kerckhoven, Wim De Meester, Gert Raskin, Stephanie De Ruyter and 
Pierre Royer  of the (former) staff of the Instituut voor Sterrenkunde of
 the KULeuven for the monitoring observations on the Euler telescope. We thank
the referee (A.A. Zijlstra) for the detailed and constructive remarks. TM
acknowledges financial support from the Fund for Scientific Research of 
Flanders (FWO) under the grant G.0178.02..
HVW acknowledges financial support from the FWO as post-doc.
The Swedish-ESO
Submillimetre Telescope, SEST, is operated jointly by ESO and the Swedish
 National Facility for Radio Astronomy, Onsala Space Observatory at Chalmers
University of Technology.

\end{acknowledgements}

\end{document}

%% file: H4172mac
\newcount\longrefs
\def\aap{\ifnum\longrefs=1 {Astron.\ Astrophys.}\else 
                           {A\hbox{\rm \&}A}\fi}
\def\aapr{\ifnum\longrefs=1 {Astron.\ Astrophys.\ Rev.}\else 
                            {A\hbox{\rm \&}AR}\fi}
\def\aaps{\ifnum\longrefs=1 {Astron.\ Astrophys.\ Suppl.}\else 
                            {A\hbox{\rm \&}AS}\fi}
\def\aj{\ifnum\longrefs=1 {Astron.\ J.}\else 
                          {AJ}\fi} 
\def\ao{\ifnum\longrefs=1 {Applied Optics}\else 
                           {Appl.\ Opt.}\fi} 
\def\aspcs{\ifnum\longrefs=1 {Astron.\ Soc.\ Pacific Conf. Series}\else 
                           {ASP Conf.\ Ser.}\fi} 
\def\apj{\ifnum\longrefs=1 {Astrophys.\ J.}\else 
                           {ApJ}\fi} 
\def\apjl{\ifnum\longrefs=1 {Astrophys.\ J. Lett.}\else 
                            {ApJ}\fi} 
\def\aplett{\ifnum\longrefs=1 {Astrophys.\ J. Lett.}\else 
                            {ApJ}\fi} 
\def\apjs{\ifnum\longrefs=1 {Astrophys.\ J. Suppl.}\else 
                            {ApJS}\fi}
\def\apss{\ifnum\longrefs=1 {Astrophys.\ and Space Science}\else 
                            {Ap\hbox{\rm \&}SS}\fi}
\def\araa{\ifnum\longrefs=1 {Ann.\ Rev.\ Astron.\ Astrophys.}\else 
                            {ARA\hbox{\rm \&}A}\fi}
\def\azh{\ifnum\longrefs=1 {Astronomicheskii Zhurnal}\else 
                            {Astron.\ Zhur.}\fi}
\def\baas{\ifnum\longrefs=1 {Bull.\ Am.\ Astron.\ Soc.}\else 
                            {BAAS}\fi}
\def\bain{\ifnum\longrefs=1 {Bull.\ Astronom.\ Institutes Netherlands}\else
                            {Bull.\ Astr.\ Inst.\ Neth.}\fi}
\def\gca{\ifnum\longrefs=1 {Geochim.\ Cosmochim.\ Acta}\else 
                           {Geochim.\ Cosmochim.\ Acta}\fi}
\def\grl{\ifnum\longrefs=1 {Geophys.\ Res.\ Lett.}\else 
                           {Geoph.\ Res.\ Lett.}\fi}
\def\iaucirc{\ifnum\longrefs=1 {IAU Circulars}\else 
                          {IAU Circ.}\fi}
\def\ip{\ifnum\longrefs=1 {in press}\else 
                          {in press}\fi}
\def\jgr{\ifnum\longrefs=1 {J.\ Geophys.\ Res.}\else 
                           {J.\ Geophys.\ Res.}\fi}  
\def\jrasc{\ifnum\longrefs=1 {J.\ Royal Astron.\ Soc.\ Canada}\else 
                           {JRAS Can.}\fi}  
\def\mnras{\ifnum\longrefs=1 {Mon.\ Not.\ Roy.\ Astron.\ Soc.}\else 
                             {MNRAS}\fi} 
\def\nat{\ifnum\longrefs=1 {Nature}\else 
                           {Nat}\fi}
\def\pasj{\ifnum\longrefs=1 {Pub.\ Astron.\ Soc.\ Japan}\else 
                            {PASJ}\fi} 
\def\pasp{\ifnum\longrefs=1 {Pub.\ Astron.\ Soc.\ Pacific}\else 
                            {PASP}\fi} 
\def\physscr{\ifnum\longrefs=1 {Physica Scripta}\else 
                            {Phys.\ Scrip.}\fi} 
\def\planss{\ifnum\longrefs=1 {Planetary \& Space Science}\else 
                            {Plan. \& Space Sci.}\fi} 
\def\procspie{\ifnum\longrefs=1 {Proc.\ SPIE}\else 
                            {Proc.\ SPIE}\fi} 
\def\qjras{\ifnum\longrefs=1 {Quarterly J.\ Royal Astron.\ Soc.}\else 
                            {QJRAS}\fi} 
\def\sa{\ifnum\longrefs=1 {Soviet Astron..}\else 
                               {Sov.\ Astron.}\fi}
\def\skytel{\ifnum\longrefs=1 {Sky \& Telescope}\else 
                            {Sky \& Tel.}\fi} 
\def\solphys{\ifnum\longrefs=1 {Solar Phys.}\else 
                               {Solar Phys.}\fi}
\def\ssr{\ifnum\longrefs=1 {Space Science Rev.}\else 
                               {Space\ Sci.\ Rev.}\fi}

\def\dutch{\def\refname{Referenties}\def\abstractname{Samenvatting}%
  \def\bibname{Bibliografie}\def\chaptername{Hoofdstuk}%
  \def\appendixname{Bijlage}\def\contentsname{Inhoudsopgave}%
  \def\listfigurename{Lijst van figuren}\def\listtablename{Lijst van tabellen}%
  \def\indexname{Index}\def\figurename{Figuur}\def\tablename{Tabel}%
  \def\partname{Deel}\def\enclname{Bijlage(n)}\def\ccname{Ter attentie van}%
  \def\headtoname{Aan}\def\headpagename{Pagina}%
  \def\today{\number\day\space\ifcase\month\or januari\or februari\or maart\or%
     april\or mei\or juni\or juli\or augustus\or september\or oktober\or%
     november\or december\fi \space\number\year}%
  \typeout{
              >>>>> use hlatex209 for Dutch hyphenation <<<<< 
         }}
\newcounter{onefig} \newcounter{fignumber}
\newcount\nocaptions \newcount\nofigures \newcount\figwidth
\newcount\viewgraphs
  \def\paper{}  \def\figlabel{} 
\long\def\nextfig#1{\setcounter{figure}{\value{fignumber}}
  \addtocounter{fignumber}{1}
  \ifnum \viewgraphs=1 \newpage \pagestyle{empty} \fi 
  \ifnum\value{onefig}=0 #1 \fi                 
  \ifnum\value{onefig}=\value{fignumber} #1 \fi}
\def\figwidths#1#2{\ifnum \nocaptions=1 #2mm \else #1mm \fi}  
\def\paper#1{}  
\long\def\plotfig#1#2{\ifnum \nofigures=1 \else #2 \fi}
\long\def\captiontext#1{\ifnum \nofigures=1 \raggedright \fi 
   \ifnum \nocaptions=1 \paper
     \ifnum \viewgraphs=0 
       \newline  \mbox{}\hrulefill\mbox{} \newline 
       \newline label:~\{\figlabel\} 
     \fi 
     \else \ifnum \nofigures=0 \fi 
   #1 \fi}

\newcount\panelwidth \newcount\panelheight 
\newcount\bxmin \newcount\bymin \newcount\bxmax \newcount\bymax
\newcount\tbxmin \newcount\tbymin
\newcount\tpanelwidth \newcount\tpanelheight \newcount\tpdif
\def\panelsize #1,#2;{\panelwidth=#1 \panelheight=#2}  
\def\setbb #1,#2;#3,#4;#5,#6;{
  \tbxmin=#1 \tbymin=#2    
  \bxmin=#3 \bymin=#4      
  \bxmax=#5 \bymax=#6}     
\def\barepanel #1{%
  \ifnum\panelheight=0 
    \tpdif=\bymax \advance\tpdif by -\bymin
    \multiply \tpdif by \panelwidth
    \tpanelheight=\tpdif
    \tpdif=\bxmax \advance\tpdif by -\bxmin
    \divide \tpanelheight by \tpdif
  \else \tpanelheight=\panelheight \fi
  \epsfig{file=#1,%
     bbllx=\bxmin bp,bblly=\bymin bp,bburx=\bxmax bp,bbury=\bymax bp,clip=,%
     width=\panelwidth mm,height=\tpanelheight mm}}
\def\labelypanel #1{
  \ifnum\panelheight=0 
    \tpdif=\bymax \advance\tpdif by -\bymin
    \multiply \tpdif by \panelwidth
    \tpanelheight=\tpdif
    \tpdif=\bxmax \advance\tpdif by -\bxmin
    \divide \tpanelheight by \tpdif
  \else \tpanelheight=\panelheight \fi
  \tpdif=\bxmax \advance\tpdif by -\tbxmin
  \tpanelwidth=\panelwidth \multiply \tpanelwidth by \tpdif
  \tpdif=\bxmax \advance\tpdif by -\bxmin
  \divide \tpanelwidth by \tpdif
  \epsfig{file=#1,%
    bbllx=\tbxmin bp,bblly=\bymin bp,bburx=\bxmax bp,bbury=\bymax bp,%
    clip=,width=\tpanelwidth mm,height=\tpanelheight mm}}
\def\labelxpanel #1{%
  \ifnum\panelheight=0 
    \tpdif=\bymax \advance\tpdif by -\bymin
    \multiply \tpdif by \panelwidth
    \tpanelheight=\tpdif
    \tpdif=\bxmax \advance\tpdif by -\bxmin
    \divide \tpanelheight by \tpdif
  \else \tpanelheight=\panelheight \fi
  \tpdif=\bymax \advance\tpdif by -\tbymin
  \multiply \tpanelheight by \tpdif
  \tpdif=\bymax \advance\tpdif by -\bymin
  \divide \tpanelheight by \tpdif
  \epsfig{file=#1,%
    bbllx=\bxmin bp,bblly=\tbymin bp,bburx=\bxmax bp,bbury=\bymax bp,%
    clip=,width=\panelwidth mm,height=\tpanelheight mm}}
\def\labelxypanel #1{%
  \ifnum\panelheight=0 
    \tpdif=\bymax \advance\tpdif by -\bymin
    \multiply \tpdif by \panelwidth
    \tpanelheight=\tpdif
    \tpdif=\bxmax \advance\tpdif by -\bxmin
    \divide \tpanelheight by \tpdif
  \else \tpanelheight=\panelheight \fi
  \tpdif=\bxmax \advance\tpdif by -\tbxmin
  \tpanelwidth=\panelwidth \multiply \tpanelwidth by \tpdif
  \tpdif=\bxmax \advance\tpdif by -\bxmin
  \divide \tpanelwidth by \tpdif 
  \tpdif=\bymax \advance\tpdif by -\tbymin 
  \multiply \tpanelheight by \tpdif
  \tpdif=\bymax \advance\tpdif by -\bymin
  \divide \tpanelheight by \tpdif
  \epsfig{file=#1,%
    bbllx=\tbxmin bp,bblly=\tbymin bp,bburx=\bxmax bp,bbury=\bymax bp,%
    clip=,width=\tpanelwidth mm,height=\tpanelheight mm}}

\def\CC{\par \vspace*{-2ex} \footnotesize \baselineskip=8pt \begin{verbatim}}

\long\def\startignore #1\stopignore{}   

\def\setlistparams{         
  \topsep=0.7ex                 
  \itemsep=0.7ex                
  \leftmargini=3ex}             
\setlistparams                  

\newcounter{alistindex}       

\newcounter{romenumnr}

\newlength{\minipagewidth}

\newsavebox{\boxcontent}
\newcommand{\ovalhead}[1]{
  \unitlength=1cm
  \sbox{\boxcontent}{\mbox{~~{#1}~~}}
  \begin{center}
    \ifdim\wd\boxcontent>6ex 
    \ifdim\wd\boxcontent<8cm 
    \begin{picture}(8,3) \thicklines     
      \put(4.0,0.8){\oval(8,1.6)} 
      \put(0.0,0.7){\parbox{8cm}{
         \begin{center} \usebox{\boxcontent} \end{center}}}
    \end{picture}
    \else \ifdim\wd\boxcontent<12cm 
    \begin{picture}(12,3) \thicklines     
        \put(6.0,0.8){\oval(12,1.6)} 
        \put(0.0,0.7){\parbox{12cm}{
           \begin{center} \usebox{\boxcontent} \end{center}}}
    \end{picture}
    \else
    \begin{picture}(16,3) \thicklines     
        \put(8.0,0.8){\oval(16,1.6)} 
        \put(0.0,0.7){\parbox{16cm}{
           \begin{center} \usebox{\boxcontent} \end{center}}}
    \end{picture}
    \fi \fi \fi
  \end{center}} 

\newcounter{headnr}            
\newcounter{subheadnr}[headnr]
\newcounter{subsubheadnr}[subheadnr]
\def\head #1\par{
  \stepcounter{headnr}                          
  \vspace{2ex} \noindent                        
  {\bf \theheadnr~~~~#1}\\[1ex] \noindent}      
\def\subhead #1\par{  
  \stepcounter{subheadnr}
  \vspace{1.3ex} \noindent
  {\bf \theheadnr.\arabic{subheadnr}~~~#1}\\[0.3ex] \noindent}
\def\subsubhead #1\par{
  \stepcounter{subsubheadnr}
  \vspace{1.0ex} \noindent
  {\bf \theheadnr.\arabic{subheadnr}.\arabic{subsubheadnr}~~~#1}\\ \noindent}

\font\dropfont= cmr12 scaled \magstep5
\def\dropcap#1#2{{\noindent
    \setbox0\hbox{\dropfont #1}\setbox1\hbox{#2}\setbox2\hbox{(}%
    \count0=\ht0\advance\count0 by\dp0\count1\baselineskip
    \advance\count0 by-\ht1\advance\count0by\ht2
    \dimen1=.5ex\advance\count0by\dimen1\divide\count0 by\count1
    \advance\count0 by1\dimen0\wd0
    \advance\dimen0 by.25em\dimen1=\ht0\advance\dimen1 by-\ht1
    \global\hangindent\dimen0\global\hangafter-\count0
    \hskip-\dimen0\setbox0\hbox to\dimen0{\raise-\dimen1\box0\hss}%
    \dp0=0in\ht0=0in\box0}#2}

\def\level #1 #2#3#4{$#1 \: ^{#2} \mbox{#3} ^{#4}$}   

\def\arcmin{\hbox{$^\prime$}}

\def\arcsec{\hbox{$^{\prime\prime}$}}

\def\kms{\hbox{km$\;$s$^{-1}$}}

\def\mic{\hbox{$\mu$m}}                     

\def\mathstacksym#1#2#3#4#5{\def#1{\mathrel{\hbox to 0pt{\lower 
    #5\hbox{#3}\hss} \raise #4\hbox{#2}}}}

\mathstacksym\lta{$<$}{$\sim$}{1.5pt}{3.5pt} 
\mathstacksym\gta{$>$}{$\sim$}{1.5pt}{3.5pt} 
\mathstacksym\lrarrow{$\leftarrow$}{$\rightarrow$}{2pt}{1pt} 
\mathstacksym\lessgreat{$>$}{$<$}{3pt}{3pt} 
